# ON PERTURBATION THEORY FOR THE STURM-LIOUVILLE PROBLEM WITH VARIABLE COEFFICIENTS

## Vladimir Kalitvianski

vladimir.kalitvianski@wanadoo.fr
CEA/Grenoble, 2009

I study different possibilities of analytically solving the Sturm-Liouville problem with variable coefficients of sufficiently arbitrary behavior with help of perturbation theory (PT). I show how the problem can be **reformulated** in order to eliminate big (or divergent) corrections. I obtain correct formulae in case of smooth as well as in case of step-wise (piece-constant) coefficients. I build simple, but very accurate analytical formulae for calculating the lowest eigenvalue and the ground state eigenfunction. I advance also new boundary conditions for obtaining more precise initial approximations. I demonstrate how one can optimize the PT calculation with choosing better initial approximations and thus diminishing the perturbative corrections. "Dressing", "Rebuilding, and "Renormalizations" are discussed in Appendices 4 and 5.

## INTRODUCTION

The Sturm-Liouville problem (SLP), understood here in a narrow sense of obtaining the eigenfunctions and eigenvalues, arises in many practical applications. Despite wide use of numerical approaches, the analytical solutions also represent a certain scientific value, especially if their physical sense is clear and the analytical formulae are simple.

In this work I study possibilities of constructing analytically the SLP perturbative solutions and analyze their accuracy. Apart from practical (numerical) applications, it is interesting to be able to understand the reasons of calculation difficulties like the matrix element divergences or the matrix element not vanishing, and to be able to find the ways of eliminating these difficulties. The consideration is made on a "physical" level of rigor for simplicity.

This study was carried out many years ago, in the early 80-ies, in the Sukhumi Institute of Physics and Technology, USSR. At that time we tried to solve a particular problem of the simplest analytical description of the heat conduction in a non uniform 1D body. My original results were developed and published later on in [1], [2], and [3]. These works were translated and published in the West, but they were not available on Internet. The purpose of the present paper is to make the main part of my original results available for each and everyone.

The most general (self-adjoint) SLP in a limited interval $[a,b]$ reads [4]:

$$\left[\frac{d}{dx}p(x)\frac{d}{dx}+\lambda \cdot r(x)-q(x)\right]\psi(x)=0,$$
$$\alpha_a\psi(a)-\psi'(a)=0, \tag{I1}$$
$$\alpha_b\psi(b)+\psi'(b)=0,$$

where $\psi(x)$ is an eigenfunction in the interval $[a,b]$ and $\lambda$ is an eigenvalue [5]. By the variable change $dy=dx/p(x)$ the differential equation can always be reduced to the similar form (**I1**), but with $p=1$ and modified $r$ and $q$. I will consider a particular case of initially $p=1$, $q=0$ so the whole coordinate dependence (or physical system non-homogeneity) will be described with $r(x)$ solely.

### 1. A perturbation theory formulation

Thus we start from the following SLP formulation:

$$\left[\frac{d^2}{dx^2}+\lambda \cdot r(x)\right]\psi(x)=0,$$
$$\alpha_a\psi(a)-\psi'(a)=0, \quad \alpha_b\psi(b)+\psi'(b)=0. \tag{1}$$

By the variable changes [4]:

$$z(x) = z_a + \int\limits_{a}^{x} \sqrt{r(x')} dx', \quad \tilde{r}(z) = r(x(z)), \quad \Phi(z) = \tilde{r}^{1/4} \psi(x(z)), \tag{2}$$

this problem can be transformed into a Schrödinger-like equation [4] (i.e., with a "potential" term $q \neq 0$):

$$\left[ \frac{d^2}{dz^2} + \lambda - U(z) \right] \Phi(z) = 0 \quad \text{with} \quad U(z) = \frac{\left( \tilde{r}^{1/4} \right)''}{\tilde{r}^{1/4}}. \tag{3}$$

These variable changes (2) look like those in a WKB approximation, but here there are no turning points since $r(x) > 0$ everywhere in our case. The "perturbation potential" operator $U(z)$ in (3) is Hermitian – it is just a function of $z$. (The prime sign $(...)'$ means here a derivative with respect to the function argument.)

When the material properties $\tilde{r}(z)$ change smoothly (slowly with $z$), the derivatives of $\tilde{r}(z)$ in $U(z)$ are "small" and one can apply the perturbation theory (PT) to calculate the eigenfunctions and eigenvalues. The corresponding formulae are well known:

$$\Phi_n^{PT} = \Phi_n^{(0)} + \sum_{m \neq n} \frac{U_{mn}}{\lambda_n^{(0)} - \lambda_m^{(0)}} \Phi_m^{(0)} + ..., \tag{4}$$

$$\lambda_n^{PT} = \lambda_n^{(0)} + U_{nn} + \sum_{m \neq n} \frac{U_{mn} U_{nm}}{\lambda_n^{(0)} - \lambda_m^{(0)}} + ..., \tag{5}$$

$$U_{mn} = \left( \Phi_m^{(0)}, U \Phi_n^{(0)} \right) = \int \Phi_m^{(0)}(z) U(z) \Phi_n^{(0)}(z) dz. \tag{6}$$

I will not consider smooth $\tilde{r}(z)$ right now. (See **Section 7** and **Appendix 1** for some examples, though.) Rather, I will go directly to a "difficult" case of a step-wise (or a piece-constant) function $\tilde{r}(z)$, for example:

$$r(x) = \begin{cases} r_1, & a \leq x \leq x_1 \\ r_2, & x_1 \leq x \leq b \end{cases} \quad \Rightarrow \quad \tilde{r}(z) = \begin{cases} r_1, & z_a \leq z \leq z_1, \quad z_1 = z(x_1) \\ r_2, & z_1 \leq z \leq z_b, \quad z_b = z(b) \end{cases}. \tag{7}$$

We will consider $\tilde{r}(z)$ as a continuous, but extremely rapidly changing function at $z = z_1$ (a "two-layer" system like, for example, (A5.11) or so). In a certain sense we may call (7) a "discontinuous function".

It is easy to see that the matrix elements (6) with (3) diverge since $U(z)$ contains the Dirac's delta-function squared. In particular, the diagonal matrix element is equal to:

$$U_{nn} = \Phi_n^{(0)}(z_1) \Phi_n^{(0)'}(z_1) \ln \sqrt{r_1 / r_2} \quad - \frac{1}{16} \int \left[ \left( \ln \tilde{r} \right)' \right]^2 \left( \Phi_n^{(0)} \right)^2 dz. \tag{8}$$

The first term in this expression vanishes progressively when $r_2 \to r_1$, but the second one remains practically infinite if $r_2 \neq r_1$ exactly. On the other hand, the problem (1) with (7) has exact and finite solutions. In particular, the exact $\lambda_n$ are finite and can be found by numerical methods from the corresponding transcendental equation. Moreover, we can expand $\lambda_n$ directly in the transcendental equation, expand in finite series in powers of $\ln \sqrt{r_1 / r_2}$ around $\lambda_n^{(0)}$ obtained from (3) with $U(z) = 0$. So two questions arise: why do these matrix element *divergences* appear in the perturbative formulation (3) – (6) and how to eliminate them?

## 1.1. Qualitative analysis of perturbation $U(z)$ in (3)

In fact, the exact problem (1) contains two different "small" parameters of expansion: the *spatial rapidity* of changing $\tilde{r}(z)$: $(dr/dz)$ or $(\Delta r/\Delta z)$, $\Delta r = r_1 - r_2 = \text{const}$, $\Delta z \to 0$; let us call it $\xi_1$, and the *relative difference* of $\tilde{r}(z)$ in the neighboring layers $\xi_2 \propto \Delta r$. Any solution of (1) (i.e., $\psi(x)$, $\lambda$, etc.) is a function of both parameters: $f(\xi_1, \xi_2)$. The problem formulation (3) is apparently better adapted for expansions in powers of $\xi_1$ when $\xi_1$ is small: $f(\xi_1, \xi_2) \approx f(0, \xi_2) + f'(0, \xi_2) \cdot \xi_1 + \dots$. The existence of finite exact solutions for a two-layer system (1), (7) means that the exact value of $f(\infty, \xi_2)$ is finite, but it cannot be obtained from its power expansion at $\xi_1 \approx 0$ since $\xi_1$ tends actually to infinity. In order to obtain an expansion in powers of an always finite parameter $\xi_2$, we have to reformulate the original problem (1) in other terms.

For that, let us note that an exact function $\psi_n(x)$ has different "*spatial frequencies*" and different *local amplitudes* in different layers, but it is an everywhere continuous function with a continuous derivative including the point $x = x_1$. The variable changes (2) catch well these properties in case of "smooth" functions $r(x)$ (it is well known from the WKB approximation applications), but (3) makes the functions $\Phi_n(z)$ "discontinuous" in case of a step-wise $r(x)$. On the other hand, the zeroth-order approximations $\Phi_n^{(0)}(z)$ are all continuous functions (see eq. (3) with $U = 0$ and Fig. 1). That means any zeroth-order function $\Phi_n^{(0)}(z)$ is "too distant" from the exact, "discontinuous" $\Phi(z)$. A "too distant" initial approximation $\Phi^{(0)}(z)$ needs "too big" perturbative addenda to correct it. In other words, to preserve the continuous character of $\psi_n(x(z)) = \Phi_n(z)/\tilde{r}^{1/4}$ the perturbation addenda, among other things, "try to build up" the step-wise factor $\tilde{r}^{1/4}$ at $\Phi_n^{(0)}(z)$ (see **Appendix 5**).

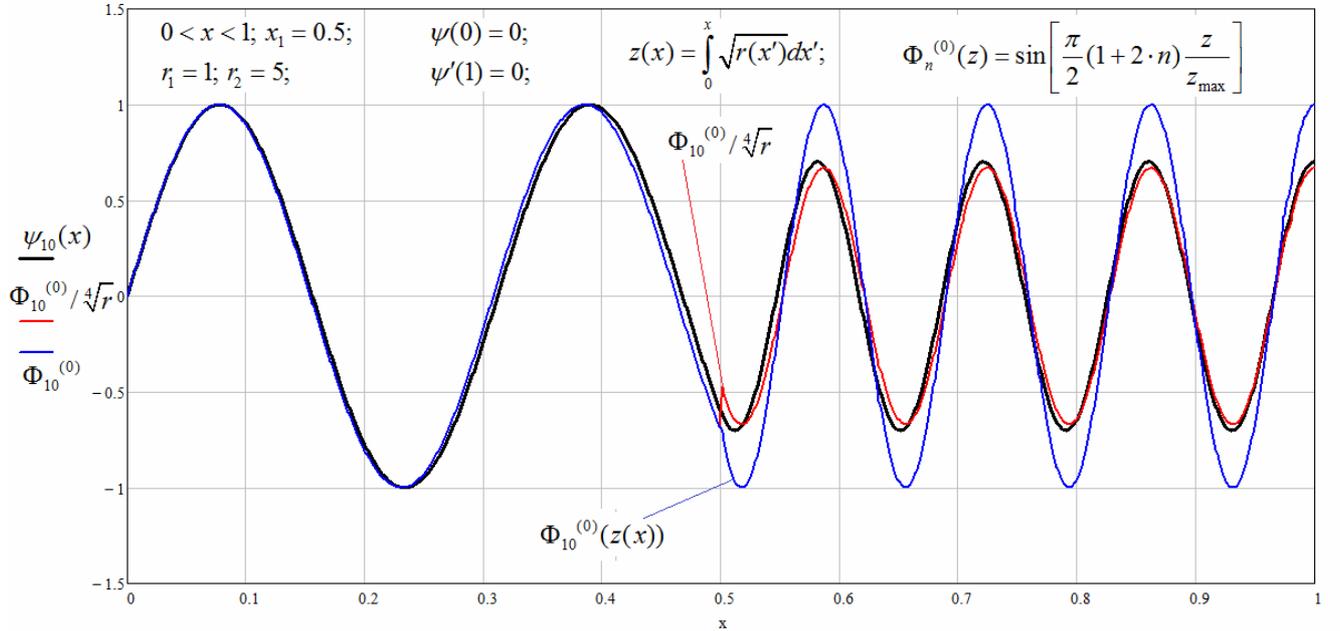

Fig.1

The exact eigenfunction $\psi_{n=10}(x)$ is an everywhere continuous function with a continuous derivative $\psi_{10}'(x)$ (the black line). The zeroth-order approximation to $\psi_{10}(x)$ (2) is discontinuous due to the factor $1/\sqrt[4]{r(x)}$ (the red line). The perturbation theory (2)-(6) must, roughly speaking, remove this "jump" and make the exact function have the right amplitudes and "spatial frequencies" everywhere. (The normalization factors are chosen to reproduce the same amplitude as $\psi_{10}(x)$ within $0 < x < x_1$. The exact function $\psi_{10}(x)$ itself was not normalized in this figure, though.)

Here it becomes clear that all spectral addenda in the exact spectral representation $\Phi_n(z) = C_n \Phi_n^{(0)}(z) + \sum_{m \neq n}^{\infty} C_m \Phi_m^{(0)}(z)$ are important and a sum truncated at finite $m \leq M$ in (4) will not be of a good accuracy. But as we know, it is the spectral coefficients $C_i$ that are expanded in the Taylor-Maclaurin series in powers of $\xi$ in the perturbation theory and it is precisely their power series diverge due to too big value of $\xi$. All this together makes "doubly truncated" PT series useless in practical calculations (unless one finds a way to formally sum up exactly all badly convergent terms).

Thus, our problem of encountering divergences in (4), (5) is mainly due to our awkward attempt to calculate finite functions $f(\infty, \xi_2)$ (i.e., $C_i(\xi_1, \xi_2)$) with help of their Taylor-Maclaurin expansion $f(\xi_1, \xi_2) \approx f(0, \xi_2) + f'(0, \xi_2) \cdot \xi_1 + ...$ obtained for formally small $\xi_1$, but used at $\xi_1 \to \infty$ (see **Appendix 5**).

## 2. Another Perturbation Theory formulation

Having understood the reason of the expansion divergence, we can make variable changes *without* this "discontinuous" factor at $\psi$ for which $\xi_1 = \infty$, for example:

$$z(x) = z_a + \int_a^x \sqrt{r(x')} dx', \quad \tilde{r}(z) = r(x(z)), \quad \varphi(z) = \psi(x(z)). \tag{9}$$

Then we obtain the following exact equation:

$$\left[ \frac{d^2}{dz^2} + \lambda - \hat{V}(z) \right] \varphi(z) = 0 \quad \text{with} \quad \hat{V}(z) = \left( \frac{1}{\sqrt{\tilde{r}}} \right)' \sqrt{\tilde{r}} \frac{d}{dz} = -\left( \ln \sqrt{\tilde{r}} \right)' \frac{d}{dz}. \tag{10}$$

If we treat $\hat{V}$ as a perturbation, then, figuratively speaking, we will take "blue lines" like in Fig. 1 as the initial approximations to the "black ones". The spatial "frequencies" and the local amplitudes are corrected in (10) exclusively with the finite perturbative terms now. Indeed, as far as the logarithm derivative is proportional maximum to the first degree of the Dirac's delta-function, the matrix elements are always *finite*. For (7) they are:

$$V_{nm} = \varphi_n^{(0)}(z_1) \cdot \varphi_m^{(0)'}(z_1) \cdot \ln \sqrt{r_1 / r_2}. \tag{11}$$

They resemble the first term from (8). If the material properties in layers are close to each other: $r_1 \approx r_2$, the matrix elements are small and give small perturbative corrections to the zeroth-order (or initial) approximations. The formal "small parameter" in (10) is now indeed a *relative difference* of $r$ in the neighbouring layers. That's it. So the original problem (1) **can be reformulated** starting from different (better) initial approximations with evidently better perturbation theory behaviour. The PT series are finite from the very beginning. The finite PT series in powers of $\xi_2$ for multi-layer systems converge fine since there are no particularities of our functions $f(\infty, \xi_2)$ at the points $\xi_2 = 0$ (see **Section 4**).

Such a situation corresponds to the Taylor-Maclaurin expansion of a finite and regular $f(\infty, \xi_2)$ in powers of $\xi_2$: $f(\infty, \xi_2) \approx f(\infty, 0) + f'(\infty, 0) \cdot \xi_2 + f''(\infty, 0) \cdot \xi_2^2 / 2! + ...$, and a truncated series can successfully be used at small $\xi_2$ values.

This is a correct scientific approach to resolving the "correction divergence" problem in case of existence of the physically reasonable exact solution. Temptation to simply discard the divergent terms in $U_{mn}$ (like discarding the second term in (8), kind of eigenvalue "renormalization") is not a scientifically justified motivation whatever "ideology" is used for that. These corrections are formally necessary, for example, to build, *after properly summing up*, the "discontinuous" exact functions $\Phi_n(z)$ from continuous functions $\Phi_m^{(0)}(z)$ (see **Appendices 4** and **5**).

### 2.1. Analysis of perturbation (10)

The perturbation operator (10) is non-Hermitian in the linear space $\{\varphi_n^{(0)}\}$ of $\hat{H}^{(0)} = \dfrac{d^2}{dz^2}$: $V_{nm} \neq V_{mn}$, but this fact is not significant for the perturbation theory applications. The only place where it comes explicitly into play is a popular statement that the second-order correction to the lowest eigenvalue $\lambda_0^{(0)}$ is always negative (the third term in the right-hand side of (5) for $n = 0$). For a non-Hermitian perturbation $V_{nm} \neq V_{mn}$ this statement does generally not hold. In this respect we have to note that the second-order correction sign characterises the property of the expanded function of being concave or convex at the expansion point. It is evident that the perturbation theory is not "obliged" to deal with convex functions $\lambda_0(\xi)$ solely (see **Appendix 2** for details and the signs at $\varepsilon^2$ in (A3.4) and in Subsection **A3.3**).

A more important finding is that the perturbation theory for (10) with (7) still needs a careful treatment that leads to another functional dependence of the small parameter in (11) due to a certain discontinuity of the exact eigenfunctions derivatives $\varphi_n'$ at $z = z_1$: the "logarithm" $\ln\sqrt{r_1/r_2}$ should be replaced with $\xi_2 = \xi_2\left(\sqrt{r_1}/\sqrt{r_2}\right) = 2\left(\sqrt{r_1/r_2}-1\right)/\left(\sqrt{r_1/r_2}+1\right)$, $|\xi_2| \leq 2$ (see **Appendix 3**). But the correct small parameter $\xi_2$ differs numerically from the "logarithm" starting only from the third order and with a small coefficient $(1/12)$, so I will keep the logarithm for my numerical examples where $\left|\ln\sqrt{r_1/r_2}\right| \leq 0.75$.

The finite PT series in powers of $\xi_2$ for multi-layer systems converge fine – their coefficients quickly decrease in the absolute value. (In fact, these *finite* series **can** even be obtained from *divergent* series (3) – (6), but with a lot of difficulties (see **Appendices 4** and **5**).)

## 3. The lowest eigenvalue $\lambda_0(\xi)$

Before passing to numerical examples for our finite PT, I would like to introduce a very accurate formula for the lowest eigenvalue $\lambda_0(\xi)$. The lowest eigenvalue for a finite physical system may help estimate the time of reaching a steady state from a transient as it determines the so called "regular regime" [2]. (For Quantum Mechanical problems the lowest eigenvalue represents the ground state energy of a compound system.)

A new formula for $\lambda_0(\xi)$ is also presented as an expansion in powers of $\xi$, but some part of its series (5) is already summed up in it into a non-trivial function of $\xi$, so the remaining perturbative corrections are much *smaller* than those in the regular PT (5). (This situation is similar to soft photon treatment in QED.)

I proceed from the following two well known facts: 1) the exact eigenvalues and the zeroth-order approximations $\lambda_n^{(0)}$ grow rapidly when $n \to \infty$: $\lambda_{n\to\infty} \propto n^2$ and 2) the relative contribution of the perturbative corrections to $\lambda_n^{(0)}$ decrease rapidly when $n \to \infty$, so the initial approximations converge quickly to the exact eigenvalues in this limit: $\lambda_{n\to\infty}^{(0)} \to \lambda_n$ (in a relative sense).

Now, the original equation (1) has a Green's function $G_\lambda(x, x')$ whose spectral representation is well known:

$$\frac{d^2}{dx^2}G_\lambda(x, y) + \lambda \cdot r(x) \cdot G_\lambda(x, y) = -\delta(x-y), \quad G_\lambda(x, y) = \sum_{n=0}^{\infty}\frac{\psi_n(x)\psi_n(y)}{\lambda_n - \lambda}. \quad (12)$$

In particular, there is a simple sum rule:

$$\int_a^b G_0(x,x)r(x)dx = \sum_{n=0}^{\infty} \frac{1}{\lambda_n} . \tag{13}$$

The Green's function $G_0(x,y)$ is exactly constructible (its equation is very simple) so this integral can be calculated *exactly*. It contains the material properties $r(x)$ in a non-trivial way via the weighting function $r(x)$ in the integrand.

If I subtract the exact sum $\sum_{n=1}^{\infty} \frac{1}{\lambda_n}$ from (13), I will obtain the exact value of $\frac{1}{\lambda_0}$. If I subtract an approximate sum $\left(\sum_{n=1}^{\infty} \frac{1}{\lambda_n}\right)^{approx.}$, I will obtain an approximate value $\left(\frac{1}{\lambda_0}\right)^{approx.}$. Now, I can use the perturbative approximate eigenvalues $\lambda_n^{PT}$ in the subtracted sum because this sum is nearly equal to the exact sum $\sum_{n=1}^{\infty} \frac{1}{\lambda_n}$ due to rapid convergence of $\lambda_n^{PT}$ to $\lambda_n$ for increasing $n$. The numerical advantage of such a calculation will be demonstrated and proven below.

The transformed equations (3) or (10) have also their own Green's function $\Gamma_\lambda(z,z')$ and in particular $\Gamma_{\lambda=0}(z,z)$. It is unknown, but like $\lambda_n$ it can be represented as a perturbation series too. Let us denote the latter as $\Gamma_0^{PT}$. Then we have symbolically:

$$\int \Gamma_0^{PT}(z,z)dz = \sum_{n=0}^{\infty} \frac{1}{\lambda_n^{PT}} \qquad \text{or} \qquad \sum_{n=1}^{\infty} \frac{1}{\lambda_n^{PT}} = \int \Gamma_0^{PT}(z,z)dz - \frac{1}{\lambda_0^{PT}} .$$

Now I write the relationship which is exact upon summing the corrections up:

$$\frac{1}{\lambda_0} = \frac{1}{\lambda_0} + \sum_{n=1}^{\infty} \left(\frac{1}{\lambda_n} - \frac{1}{\lambda_n^{PT}}\right) = \frac{1}{\lambda_0^{PT}} + \left(\int_a^b G_0(x,x)r(x)dx - \int \Gamma_0^{PT}(z,z)dz\right). \tag{14}$$

Indeed, if one manages to sum up all powers of perturbative corrections, one obtains the identities: $\left(\lambda_n^{PT}\right)_{summed} = \lambda_n$, $\left(\Gamma_0^{PT}(z,z)\right)_{summed} = \Gamma_0(z,z)$, and the round bracket with Green's functions disappears. If one uses the PT series truncated in practice at some finite order, one obtains an approximate value of $\lambda_0^{-1}$ or $\lambda_0$, which I denote hereafter as $\lambda_0^{GF}$ ("GF" stands for Green's Function sum rule).

I rewrite (14) in another symbolical way:

$$\lambda_0^{GF} = \left\{\frac{1}{\lambda_0^{PT}} + \left(\int_a^b G_0(x,x)r(x)dx - \int \Gamma_0^{PT}(z,z)dz\right)\right\}^{-1} . \tag{15}$$

This formula expresses the lowest eigenvalue via perturbative series in powers of the small parameter $\xi$, and the question arises: why and how is it different from the pure perturbative expansion of $\lambda_0^{PT}$? The answers are that a part of perturbation series in it is summed up exactly into a non trivial function of $r(x)$, for example, of $r_1$ and $r_2$ for a two layer system, thanks to the *exact* Green's function sum rule, thus the remaining series is different. Indeed, the zeroth-order approximation $\lambda_0^{GF(0)}(\xi) = \left\{\frac{1}{\lambda_0^{(0)}} + \left(\int_a^b G_0(x,x)r(x)dx - \int \Gamma_0^{(0)}(z,z)dz\right)\right\}^{-1}$

contains already $r_1 \neq r_2$ due to the first integral calculated *exactly* and $\lambda_0^{GF(0)}(\xi)$ it is obviously different from $\lambda_0^{(0)} = \lambda_0(\xi = 0)$ where $r_1 = r_2$ (compare formulae (17) and (18)). If we expand the first integral in (15) in powers of $\xi$ to the same order as the second integral, the integrals with the Green's functions will cancel and we return to the ordinary PT expansion: $\lambda_0^{GF} = \lambda_0^{PT}$.

It is easy to prove analytically that the "GF" formula (15) is more accurate than the perturbation series formula $\lambda_0^{PT} = \lambda_0^{(0)} + V_{00} + \dots$. For that let us denote the relative error of $\lambda_n^{PT}$ as $\delta_n^{PT} = \left(\lambda_n^{PT} - \lambda_n\right)/\lambda_n^{PT}$. Then the formula's (15) relative error $\delta_0^{GF} = \left(\lambda_n^{GF} - \lambda_n\right)/\lambda_n^{GF}$ is expressed in the following way:

$$\delta_0^{GF} = -\sum_{n=1}^{\infty} \frac{\lambda_0}{\lambda_n} \cdot \delta_n^{PT}. \tag{16}$$

To estimate the latter sum, let us replace the ratio $\lambda_0 / \lambda_n$ with $(n+1)^{-2}$ and replace all decreasing $\left|\delta_n^{PT}\right|$ with a constant (maximum) relative error, for example, with $\left|\delta_1^{PT}\right|$. Then $\left|\delta_0^{GF}\right| \leq (2/3)\left|\delta_1^{PT}\right|$ in the same PT order. We know that in fact $\delta_n^{PT} \to 0$ when $n \to \infty$. In particular, $\left|\delta_0^{PT}\right| < \left|\delta_1^{PT}\right|$ so normally $\left|\delta_0^{GF}\right| << \left|\delta_0^{PT}\right|$. Actually, $\left|\delta_0^{GF}\right|$ is even smaller than $(2/3)\left|\delta_1^{PT}\right|$ since $\delta_n^{PT}$ are decreasing in the absolute values, but they may also have different signs in (16) and thus partially cancel each other. In other words, the initial approximation $\left(\lambda_0^{GF}\right)^{(0)}(\xi)$ in formula (15) is much *closer* to the exact value and thus the perturbative corrections to it are much *smaller*. The numerical demonstrations including the formula (15) accuracy are given in Fig. 2 – Fig. 6.

## 4. Numerical examples

For a numerical example let us consider a two-layer system with the normalized interval $z \in [0,1]$ and the boundary conditions: $\varphi(0) = 0$, $\varphi'(1) = 0$. I vary the first layer effective thickness within $0 \leq z_1 \leq 1$ and the second one varies then also within $0 \leq (1 - z_1) \leq 1$. The eigenvalue expansions are made in powers of $\varepsilon = \ln(r_1 / r_2)$. I vary it within $\varepsilon \in [-1.5, 1.5] \Rightarrow 4.5^{-1} \leq r_1 / r_2 \leq 4.5$, so the material property difference may be not so small.

The transcendental equation for $\lambda$ is: $e^{\varepsilon/2} \cdot ctg(\sqrt{\lambda} \cdot z_1) = tg\left[\sqrt{\lambda} \cdot (1 - z_1)\right]$, $\lambda_0^{PT(0)} = \dfrac{\pi^2}{4}$.

The first-order PT and GF approximations for the lowest eigenvalue are the following:

$$\lambda_0^{PT(1)} = \frac{\pi^2}{4}\left[1 + \frac{\overbrace{\frac{4}{\pi^2}V_{00}}}{\frac{\varepsilon}{\pi}\sin(\pi z_1)}\right], \tag{17}$$

$$\lambda_0^{GF(1)} = \left\{ \underbrace{\left(\lambda_0^{(0)}\right)^{-1}}_{\frac{4}{\pi^2}}\left[1 - \overbrace{\frac{\frac{4}{\pi^2}V_{00}}{\frac{\varepsilon}{\pi}\sin(\pi z_1)}}\right] + \overbrace{\frac{\int G_0(x,x)r(x)dx}{z_1(1-z_1)e^{-\varepsilon/2}}} - \overbrace{\frac{\int \Gamma_0^{PT(1)}(z,z)dz}{z_1(1-z_1)\left(1 - \frac{\varepsilon}{2}\right)}}\right\}^{-1}. \tag{18}$$

When $\varepsilon \to 0$, the perturbative corrections to $\lambda_0^{(0)}$ tend to zero both in (17) and (18).

When $z_1$ approaches 0 or 1 (one of two "effective" layers is made thin), then the perturbative corrections must tend to zero too. I did not mention this fact in the main text, but the thickness of a layer may also be a natural small parameter. The perturbation (10) catches this fact (see **Section 5**, though). In order to verify the precision of (17) and (18), I plotted Fig. 2 – Fig.5 where I made calculations at the following discrete points:

$$\varepsilon_i = -1.5 + 3(i-1)/20, \quad 1 \le i \le 21, \qquad (z_1)_k = 0.05(k-1), \quad 1 \le k \le 21,$$

One can see that the GF zeroth-order approximation $\lambda_0^{GF(0)}$ is better than the PT first-order approximation $\lambda_0^{PT(1)}$, and the first-order approximation $\lambda_0^{GF(1)}$ is very accurate in the considered 2D region.

In case of equal "effective" layer thicknesses ($z_1 = 1/2$), the small parameter $\varepsilon$ contribution is not suppressed with a small factor due to a small layer thickness. In this case the transcendental equation has the exact solutions, for example: $\lambda_0|_{(z_1=0.5)} = \left\{ 2 \cdot \text{arctg}\left[ (r_1/r_2)^{1/4} \right] \right\}^2 = \left[ 2 \cdot \text{arctg}\left( e^{\varepsilon/4} \right) \right]^2$. The numerical accuracy of formula (18) is better than 0.5% in the region $4.5^{-1} \le r_1/r_2 \le 4.5$ (Fig. 6) and is comparable with that of $\lambda_0^{PT(3)}$ (Fig. 14).

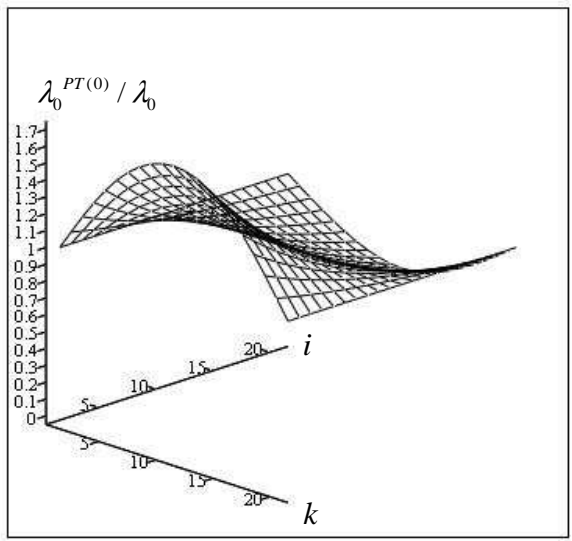

Fig. 2. Ratio of the zeroth-order PT eigenvalue to the exact one.

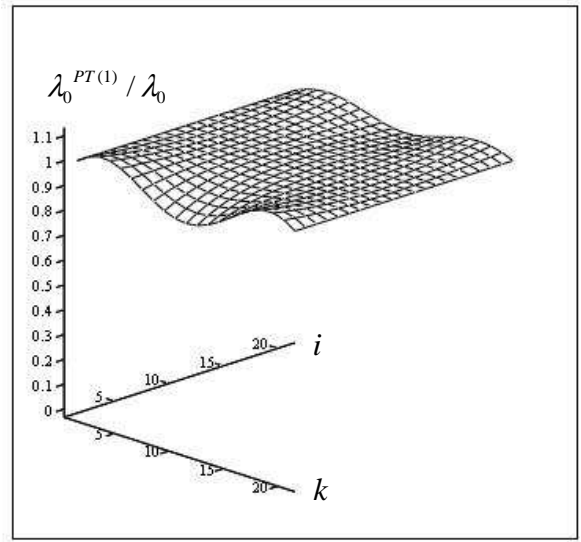

Fig. 3. Ratio of the first-order PT eigenvalue to the exact one.

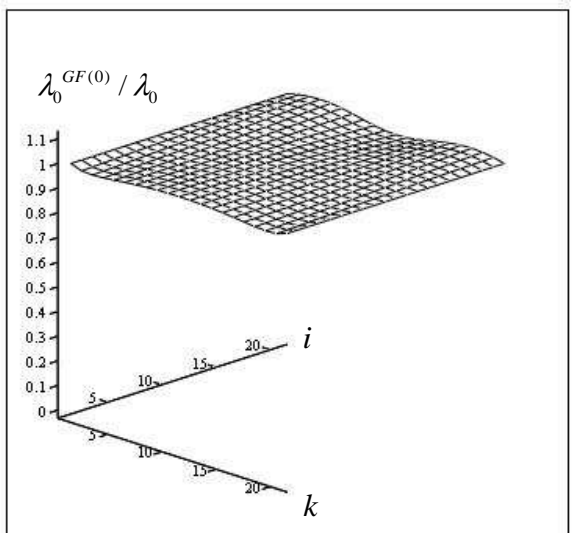

Fig. 4. Ratio of the zeroth-order GF eigenvalue to the exact one.

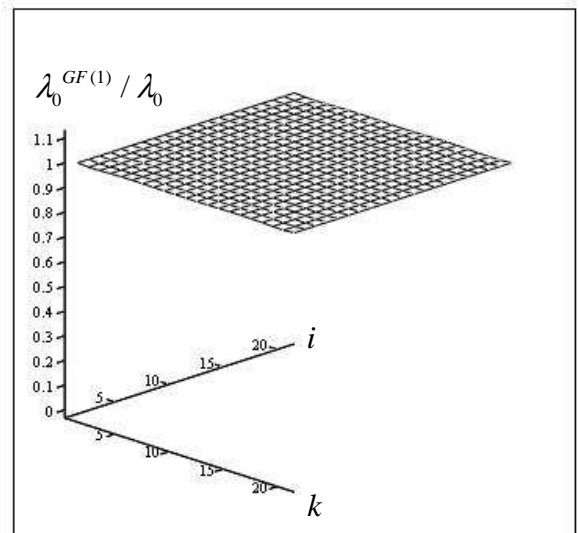

Fig. 5. Ratio of the first-order GF eigenvalue to the exact one.

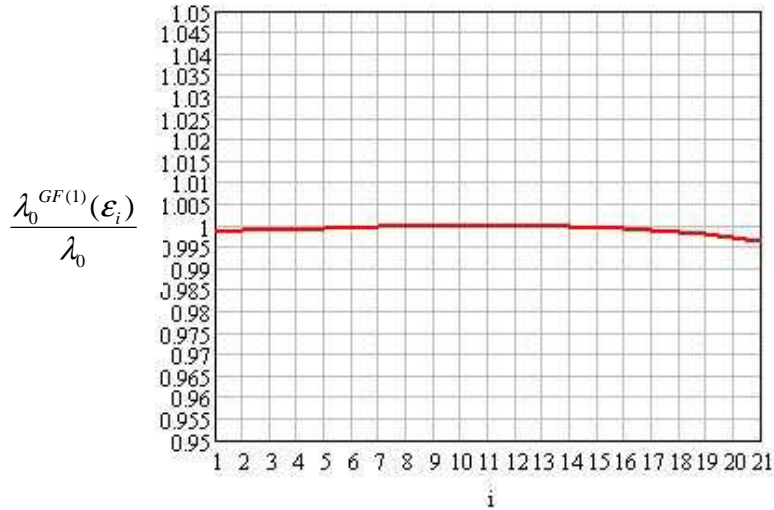

Fig. 6. Ratio of the first-order GF eigenvalue to the exact one in case of equal effective layer thicknesses $z_1 = 0.5$.

## 5. Harmonized (or well-balanced) boundary conditions

It follows from formulae (17) – (18) that when the layer thickness tends to zero, its influence (contribution) diminishes. It is natural: for example, an infinitesimally thin layer in the middle of the system, like shown in (Fig. 7), or at the system extremities (Fig. 8, Fig. 9), cannot physically change the exact solution and the zeroth-order solution is nearly exact. For an "interior" thin layer this is easily seen from the matrix element $V_{nm}$ canceling: the contributions of two $r$-jumps in $V_{nm}$ are nearly equal, but opposite in sign (see (11)).

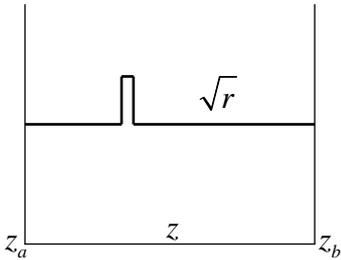

Fig. 7. A thin third layer in the middle.

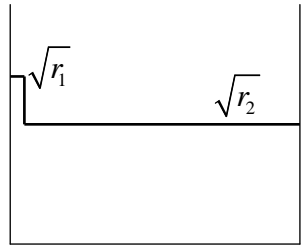

Fig. 8. A thin layer at the left end.

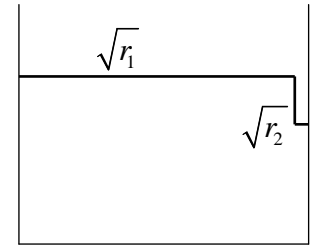

Fig. 9. A thin layer at the right end.

For the boundary conditions $\varphi(z_a) = 0$, $\varphi'(z_b) = 0$, $\varphi'(z_a) = 0$, $\varphi(z_b) = 0$, or $\varphi(z_a) = 0$, $\varphi(z_b) = 0$ this property is also automatically implemented in the matrix elements (11).

However for the general (mixed) boundary conditions (1) this is not the case. The transformed boundary conditions (1) now read:

$$\alpha_a \varphi(z_a) - \sqrt{r_1} \cdot \varphi'(z_a) = 0, \ \alpha_b \varphi(z_b) + \sqrt{r_2} \cdot \varphi'(z_b) = 0. \tag{19}$$

They determine the zeroth-order eigenfunctions $\varphi^{(0)}(z)$ and depend on the function $\tilde{r}(z)$ values at the interval extremities. When, for example, $z_1 \to z_a$ (Fig. 8), the approximate eigenfunctions $\varphi^{(0)}(z)$ and/or their derivatives $\varphi^{(0)'}(z)$ do not tend to zero because of (19) and because of independence of the first boundary condition from $z_1$. An infinitesimal first layer, inessential for the exact solution, modifies the boundary condition $\alpha_a \varphi(z_a) - \sqrt{r_1} \cdot \varphi'(z_a)$ and makes the zeroth approximation $\varphi^{(0)}(z)$ strongly dependent on $r_1$ which is not good. The non-vanishing matrix elements (11) serve factually to remove this dependence and to build finally the $r_1$-independent exact solution $\varphi(z)$.

The same situation takes place when $z_1 \to z_b$ (Fig. 9): the second infinitesimal layer, due to the variable change (9), gets involved in the second boundary condition $\alpha_b \varphi(z_b) + \sqrt{r_2} \cdot \varphi'(z_b) = 0$ and the non-zero matrix elements serve to remove this $r_2$-dependence to obtain the $r_2$-independent exact solution $\varphi(z)$. This shows once more that the perturbative correction numerical values are determined with the initial approximation quality (i.e., with its "closeness" to the exact solution).

Knowing that an infinitesimal layer cannot change the exact solution, but can change the initial approximation, we can replace the systems like in Fig. 8 and Fig. 9 by physically equivalent systems like in Fig. 10 and Fig. 11, i.e., with *additional* infinitesimal boundary layers with average characteristics. This replacing will change the boundary conditions (19) for the following ones:

$$\alpha_a \varphi(z_a) - \left\langle \sqrt{\tilde{r}} \right\rangle \cdot \varphi'(z_a) = 0, \text{ or } \alpha_b \varphi(z_b) + \left\langle \sqrt{\tilde{r}} \right\rangle \cdot \varphi'(z_b) = 0 \ , \tag{20}$$

and the matrix elements (11) will obtain the following addenda:

$$\delta V_{nm} = \varphi_n^{(0)}(z_a) \cdot \varphi_m^{(0)\prime}(z_a) \cdot \xi_2 \left( \left\langle \sqrt{r} \right\rangle / \sqrt{r_1} \right), \text{ or } \varphi_n^{(0)}(z_b) \cdot \varphi_m^{(0)\prime}(z_b) \cdot \xi_2 \left( \sqrt{r_2} / \left\langle \sqrt{r} \right\rangle \right), \tag{21}$$

where $\left\langle \sqrt{r} \right\rangle$ is some average value, for example:

$$\left\langle \sqrt{r} \right\rangle = \int\limits_{z_a}^{z_b} \sqrt{\tilde{r}(z)} dz \, / \left( z_b - z_a \right) \tag{22}$$

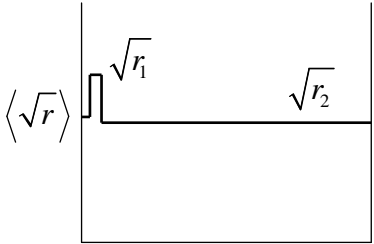

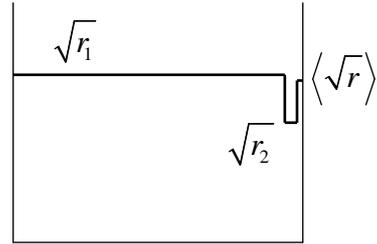

Fig. 10. An additional infinitesimal layer at the left end.　　　　Fig. 11. An additional infinitesimal layer at the right end.

Now when the first layer gets thin, the average value $\left\langle \sqrt{r} \right\rangle$ tends to $\sqrt{r_2}$, the matrix elements (11) with (21) cancel, and the zeroth-order approximation $\varphi^{(0)}(z)$ tends to the exact solution that only depends on $\sqrt{r_2}$. Similar cancellation happens when the second layer gets thin.

Thus, doing so, we do not change the exact solution, but improve the zeroth-order approximation $\varphi^{(0)}(z)$ and therefore diminish the perturbative corrections $\overline{V}_{nm}$. This is one more example of better choosing the initial approximation for the perturbation theory.

Such a system replacing may be called "harmonization" of the boundary conditions in the perturbative approach. I do not give here numerical examples of this "harmonization" since I never used it in my practice. It may well happen that the proposed choice of $\left\langle \sqrt{r} \right\rangle$ is not so optimal for numerical calculations and one may find something better.

### 6. The "ground state" eigenfunction $\psi_0$

It seemed to me that (already normalized) the "ground state" eigenfunction $\psi_0$ or $\varphi_0$ could also be calculated with a great precision with an **analogous** to (15) formula:

$$\psi_0^{GF}(x) = \sqrt{\lambda_0^{GF}} \left\{ \frac{\left(\overline{\varphi}_0^{PT}(z(x))\right)^2}{\lambda_0^{PT}} + \left[G_0(x,x) - \Gamma_0^{PT}(z(x), z(x))\right] \right\}^{1/2} . \qquad (23a)$$

However my recent quick analysis (March 2018) forced me to change my mind because there are some complications in (23a). Indeed, as far as the exact eigenfunctions $\varphi_n(z)$ are normalized with the weight $\sqrt{\tilde{r}(z)}$: $(\varphi_n, \varphi_m) = \int \sqrt{\tilde{r}(z)} \cdot \varphi_n(z) \cdot \varphi_m(z) dz = \delta_{nm}$, the zeroth-order approximations $\varphi_n^{(0)}(z)$ **must** be normalized with **some constant weight**, say, with $\langle \sqrt{r} \rangle$ from (22) rather than with 1. Let us denote them as $\overline{\varphi}_n^{(0)}(z)$. If $r = const$, the approximate and the exact eigenfunctions will coincide since their normalization conditions are now similar and $\langle \sqrt{r} \rangle = \sqrt{r} = const$ in this case. It is important, for example, in simple estimations of $\overline{\varphi}_n^{(0)}(z)$, of $\psi_n^{GF(0)}$ from (23b), and in other formulae containing the *explicit* functions $\overline{\varphi}_n^{(0)}(z)$. This necessity was already underlined in [1–3], but without numerical examples. My recent numerical calculations for the SLP with $\varphi(0) = 0$, $\varphi'(1) = 0$ showed, however, that it is not a sufficient measure, at least for the simplest $\varphi_0^{GF(0)}$. As the Green's function difference in (23a) is linear in $x$ ($G_0(x,x) = x$) and $\left(\overline{\varphi}_n^{PT(0)}\right)^2$ is quadratic in $x$ at small values of $x$, the function $\varphi_0^{GF(0)}$ (23a) behaves as a square root of $x$ at small $x$ instead of being a linear function. This surprising and "unpleasant" GF-property was only figured out by me while preparing some pictures of the simplest $\varphi_0^{GF(0)}$ in order to determine the regions where one obtains negative numbers under the square root. I did not expect a linear behaviour of the sum of $x^2$ in $\sum_{n=0}^{\infty} \frac{\psi_n^2(x)}{\lambda_n}$ (12) at small $x$ and numerically $\varphi_0^{GF(0)}$ (23a) turned out to be of a much worse accuracy than $\lambda_0^{GF(0)}$ (see Fig. 4). I should have tested this formula before proposing it simply **by analogy** with a well verified formula (18). I can say nothing about the first order version of (23a) – whether it becomes better in this respect or not, since I never constructed and analyzed it. But there are other, reliable, ways of obtaining the eigenfucntions, for example, with the regular PT formulae (4) with (11) or with $\overline{V}_{nm}$.

Apart from pure PT, for a multilayer physical system like (7) one can always write down a formal solution $\psi_\lambda(x)$, i.e., a formula parametrically dependent on $\lambda$. Injecting approximate values $\lambda_0^{GF}$, $\lambda_{n>0}^{PT}$ in $\psi_\lambda(x)$ gives approximate eigenfunctions $\psi_n(x)$ of a good accuracy. This original approach was already described in [1–3] (I tested it, but not published figures for being short).

Also one may, for the numerical purposes, replace the "discontinuous" $r(x)$ (7) with a continuous one (relatively small, but finite $\Delta x$ or $\Delta z$ in (A5.11)) for which the WKB-like solutions $\Phi_n^{(0)}(2)$ are rather precise.

Concerning the "sum rule" approach like (15) and (23a), it might happen that the following formula, *linear at small $x$*, would be a better estimate of (not yet normalized) $\psi_0^{GF}$:

$$\psi_0^{GF}(x) = Const \cdot \left\{ \frac{\overline{\varphi}_0^{PT}(z(x)) \cdot \overline{\varphi}_0^{PT}(z(x_c))}{\lambda_0^{PT}} + \left[G_0(x, x_c) - \Gamma_0^{PT}(z(x), z(x_c))\right] \right\} , \qquad (23b)$$

where $x_c$ is a fixed point within the interval $(x_a, x_b)$, for example, $x_b$ for the case $\varphi(0) = 0$, $\varphi'(1) = 0$. This and other possible "sum-rule-based" formulae are still to be analyzed and tested.

Finally, I have just tried one more approach to build the ground state eigenfunction with help of $G_0(x, y)$ and $\psi_n^{(0)}(x) = \Phi_n^{(0)}(z(x)) / \sqrt[4]{r(x)}$ (2) – (4): as far as the WKB-like functions $\psi_n^{(0)}$ reproduce well the spatial frequencies and *amplitudes* of the exact ones $\psi_n(x)$ (see Fig. 1), then they might be "well orthonormal" to the exact ones, namely, if $\psi_0^{(0)}(x)$ is "sufficiently orthogonal" to all $\psi_n(x)$, $n > 0$, then an approximate $\psi_0^{GF}(x)$ can be constructed from the exact spectral sum rule for $G_0(x, y)$ (see also **A3.2**): $\psi_0(x) / \lambda_0 = \left(\psi_0(y), G_0(x, y)\right)$, where the bracket $(\phi, \eta)$ denotes the scalar product with weight $r$. Namely:

$$\psi_0^{GF(0)}(x) = \lambda_0^{GF(1)} \cdot \int_a^b r(y) \cdot \psi_n^{(0)}(y) \cdot G_0(x, y) dy . \tag{23c}$$

Here I use a very precise approximation $\lambda_0^{GF(1)}$ (18) instead of the unknown exact value of $\lambda_0$. This formula (see analytical expression (A3.5)) works fine: figures below show a rather good accuracy of a two-layer physical system: the black (exact) and the blue (approximated with (23c)) lines are sufficiently close to each other.

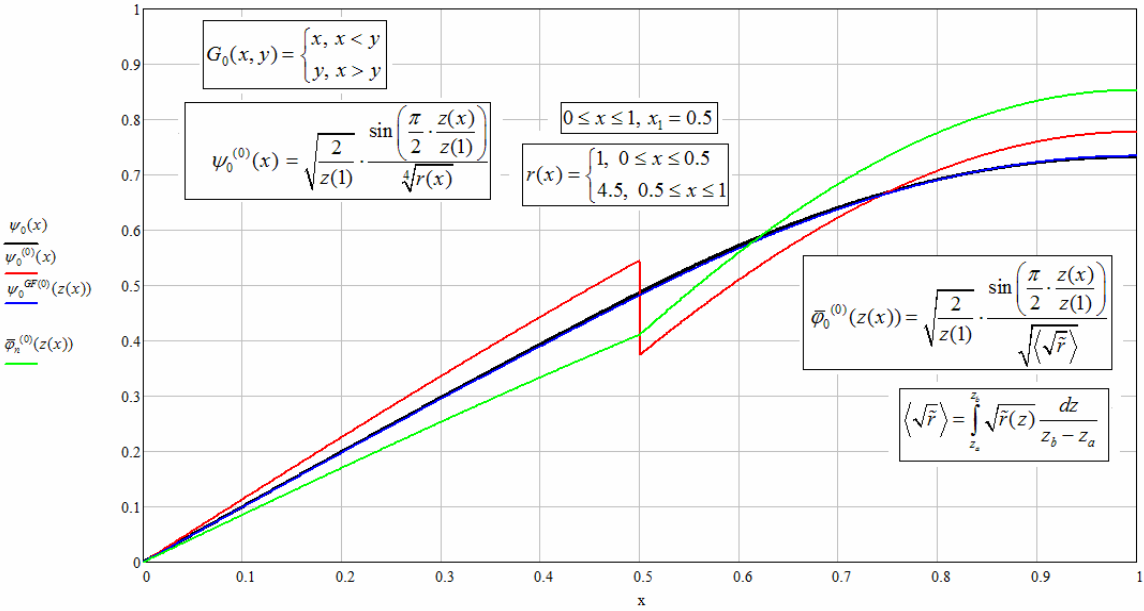

Fig. 12. The exact normalized eigenfunction $\psi_0(x)$ (black line) and approximate ones: $\psi_0^{(0)}$, $\overline{\varphi}_0^{(0)}$, and $\psi_0^{GF(0)}$ (23c) for a big difference between $r_1$ and $r_2$ (1 versus 4.5). (See **Appendix 6** for more details.)

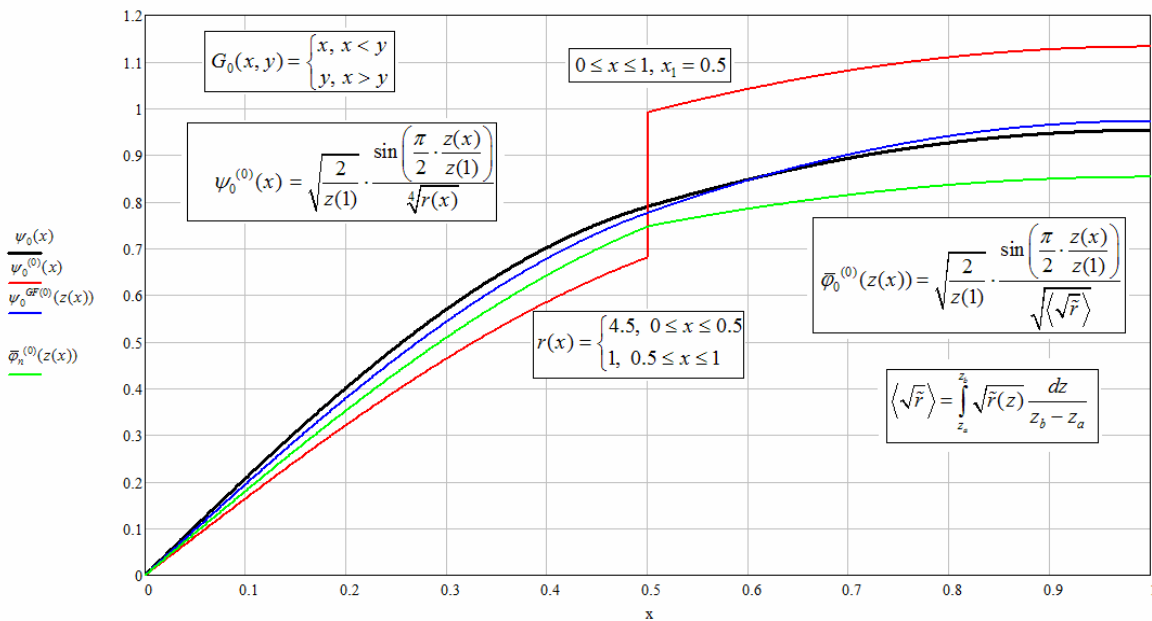

Fig. 13. The exact normalized eigenfunction $\psi_0(x)$ (black line) and approximate ones: $\psi_0^{(0)}$, $\overline{\varphi}_0^{(0)}$, and $\psi_0^{GF(0)}$ (23c) for a big difference between $r_1$ and $r_2$ (4.5 versus 1). (See **Appendix 6** for more details.)

### 7. Non plane geometry

In the examples above one could choose the lowest value of $z$ to be zero: $z_a = 0$. For a plane geometry it changes nothing. However for a hollow system with cylindrical or spherical geometries it is better not to put $z_a$ to zero. To explain why it is so, let us consider a simple uniform ($r = \text{const}$) cylindrical system:

$$\left( \frac{1}{R} \frac{d}{dR} R \frac{d}{dR} + \lambda \cdot r \right) \psi(R) = 0,$$

$$\psi(R_{\min}) = 0, \quad \psi(R_{\max}) = 0. \tag{24}$$

The eigenvalues are dependent on the "cylindricity" parameter $R_{\max} / R_{\min}$. If it is close to unity, $R_{\max} / R_{\min} \to 1$, the system is physically close to a plane one and $\lambda_n \approx \frac{\pi^2 (n+1)^2}{\Delta R^2}$, where $\Delta R \ll R_{\max}$ is the cylinder thickness. If the "cylindricity" tends to infinity: $R_{\max} / R_{\min} \to \infty$ ($R_{\min} \ll R_{\max}$), the eigenvalues have coefficients different from $\pi^2$, for example, $\lambda_0 = \frac{2.4048^2}{R_{\max}^2}$.

After the variable changes (9) there will appear an additional "perturbation" operator $\hat{V}_{cyl}(Z, r)$:

$$\left( \frac{1}{Z} \frac{d}{dZ} Z \frac{d}{dZ} + \lambda - \hat{V}_{cyl} \right) \varphi(Z) = 0, \quad \hat{V}_{cyl}(Z, r) = \frac{R(Z) - Z / \sqrt{r}}{RZ} \frac{d}{dZ}. \tag{25}$$

The zeroth-order eigenfunctions and eigenvalues depend on the new "cylindricity" parameter $Z_{\max} / Z_{\min}$. If it is not numerically equal to the original parameter $R_{\max} / R_{\min}$, the contribution of $\hat{V}_{cyl}$ will be different from zero to "restore" the right value of the exact solution. If one chooses $Z_{\min} = \sqrt{r} \cdot R_{\min}$, then $R(Z) - Z / \sqrt{r} \equiv 0$ and no contribution from $\hat{V}_{cyl}$ arises at all. It obviously simplifies calculations.

In multi-layer cylindrical systems the contribution of $\hat{V}_{cyl}$ is always different from zero, but one can optimize the choice of $Z_{\min}$ in order to minimize the $\hat{V}_{cyl}$ contribution and therefore to improve the zeroth-order approximations. Then one may even neglect the contribution of $\hat{V}_{cyl}$ in low PT orders.

Similar optimization is possible in case of the spherical geometry. The detailed formulae are given in my publications [1–3]. However, I never tested these optimizations, so they must be taken with a caution and they may admit (or need) some possible improvements.

It is interesting that equation (24) can itself be considered as a plane geometry equation with *smooth* R-dependent coefficients $p(R)$ and $r(R)$ (see eq. (1)). Then one can make the variable changes (2): $z = R - R_{\min}$, $\Phi(z) = \sqrt{z + R_{\min}} \cdot \psi(R(z))$, $U(z) = -0.25 / (z + R_{\min})^2$. Here $z_{\min} = 0$ on purpose. The first-order corrections are small if $R_{\max} \approx R_{\min}$ and they become maximal when $R_{\max} / R_{\min} \to \infty$. In this case the usual PT (5) gives: $\lambda_0^{PT(0)} \approx \frac{\pi^2}{R_{\max}^2}$, $\lambda_0^{PT(1)} \approx \frac{2.7642^2}{R_{\max}^2}$, $\lambda_0^{PT(2)} \approx \frac{2.68^2}{R_{\max}^2}$. It demonstrates a little bit slow convergence to the exact eigenvalue $\lambda_0 = \frac{2.4048^2}{R_{\max}^2}$. The non-Hermitian perturbation from (10) provides a better convergence for the lowest eigenvalue: $\lambda_0^{PT(0)} \approx \frac{\pi^2}{R_{\max}^2}$, $\lambda_0^{PT(1)} \approx \frac{2.3269^2}{R_{\max}^2}$, $\lambda_0^{PT(2)} \approx \frac{2.4035^2}{R_{\max}^2}$.

The zeroth-order GF-approximation (15) gives a quite accurate value in this limit:

$$\lambda_0^{GF(0)} \approx \frac{2.3271^2}{R_{max}^2} \, .$$

## 8. DISCUSSIONS

A brief résumé is given in the "read before reading" section – in the **Abstract**. Here I would like to underline that encountering divergent (or big) corrections is not fatal. In Theoretical Physics it may be connected with bad understanding of physical phenomena and thus with bad initial approximations and wrongly guessed interaction (perturbation) Hamiltonians. In physics we often start from equations like (3) rather than from (1) (for example, in CED with a singular self-induction due to the self-interaction guess [6]) and we puzzle where do these singularities come from? Concentrating too much on analysis of the perturbation term like (8), one can conclude that the divergence appears due to "too strong interaction at short (close to $z_1$) distances", for example. It does not advance our understanding and the "renormalization prescription" does not work in all cases without fail. A more profound and comprehensive analysis should include not only narrow matrix element behavior observations, but also general physical and mathematical reasoning; and a better problem formulation may follow just from good sense [7], [8]. As we could see from this paper, there may be many different approximate analytical expressions for the same exact variable and **we are not bound to stick to one**, especially if it is bad.

This paper deals mostly with constructing **simple analytical approximations** for the solutions as it was conceived at the times of big and expensive computers with punched card input ("IBM machines") and small scientific calculators with simple programmed functions like $\sin$, $\cos$, $\sqrt{\phantom{x}}$, $x^y$, $\ln$, etc. Nowadays many problems can be solved numerically with modern (super) computers, **but no computer can replace a creative researcher in correctly setting up a problem to solve**. For example, most "fundamental" QFT are still considered ill-defined [9] or even "non existent" [10] due to being "incomplete".

## 9. CONCLUSIONS

In the present paper I showed how important is to choose an appropriate initial approximation for building a reasonable perturbation theory. Although banal mathematically, this understanding is not widely appreciated in physics due to historical and some other (sociological) reasons. In particular, the renormalizations (modifications of bad solutions) have been given such a "state of the art" that it is extremely difficult to get through the common opinion today. In this article I demonstrate that one *can* find a **short-cut** to convergent series without appealing to "renormalizations". This short-cut consists in **reformulation** of the problems to solve in better terms and variables, which better catch the main features of the exact solution qualitatively and thus quantitatively. This possibility should be taken seriously by the scientific community.

I believe that we can reformulate QED (compare perceptions in [8] and [10]) and some other "gauge" theories in order to directly obtain their final results without renormalizations.

**APPENDIX 1**

As I mentioned in the main text, the variable changes (2) are good for smooth coefficients $r(x)$. In fact, there is a three-parametric family of coefficients $r(x)$ that leads to a constant perturbation $U(z)$:

$$r(x \mid c, d_1, d_2) = c \cdot (x - d_1)^{-2}(x - d_2)^{-2}, \; d_{1,2} \notin [a, b], \qquad (A1.1)$$

$$U = (d_1 - d_2)^2 / (4c) = const. \qquad (A1.2)$$

A constant perturbation displaces uniformly the whole "non-perturbed" spectrum $\lambda_n = \lambda_n^{(0)} + U$ and does not change the "non-perturbed" (zeroth-order) eigenfunctions $\Phi_n(z) = \Phi_n^{(0)}(z)$ because $U_{nn} = U$ and $U_{nm} = 0$, $(n \neq m)$. So the perturbation theory is developed actually "around" such a family. Functions (A1.1) are not so "smooth". This fact may be used in practical applications in order to approximate the continuous $r(x)$ with help of some $r(x \mid c, d_1, d_2)$ and obtain simple analytical solutions. (Of course, there are other particular SLP coefficients $p(x)$, $r(x)$, $q(x)$ in (I1) for which the exact analytical solutions exist.)

If $r(x)$ is "distant" from any $r(x \mid c, d_1, d_2)$, i.e., if the perturbative corrections $U_{nm}$ are relatively big, it is worth to consider another variant of PT, for example, (9) – (10). The PT corrections to the eigenfunctions (non diagonal matrix elements) demonstrate which variant of PT gives better initial approximation.

**APPENDIX 2**

As I mentioned in the main text, the perturbation operator (10) is non-Hermitian: $V_{nm} \neq V_{nm}$. The total differential operator $\hat{H}(z) = \left[ d^2 / dz^2 - \hat{V}(z) \right]$ is non-Hermitian either in the linear space of eigenfunctions of the "non-perturbed" problem (in the space of the zeroth-order eigenfunctions) unlike the operator $\hat{H}_0 = d^2 / dz^2$. The total differential operator $\hat{H}(z)$ is Hermitian in another linear space $\{\varphi_n\}$, which is determined with the following scalar product: $(\varphi_n, \varphi_m) = \int \sqrt{\tilde{r}} \cdot \varphi_n \varphi_m \, dz = \delta_{nm}$ (where $\hat{H}_0$ is non-Hermitian). This property follows from (1), (9 – 10), of course. And the original SL operator (I1) or (1) is Hermitian in the space of its eigenfunctions $\{\psi_n\}$ with the following scalar product: $(\psi_n, \psi_m) = \int r(x) \cdot \psi_n(x) \psi_m(x) \, dx = \delta_{nm}$.

**APPENDIX 3**

**A3.1. Spectral analogues of the matrix elements**

As I mentioned in the main text, the correct small parameter in the perturbative expansion with (11) is $\xi_2 = \xi_2 \left( \sqrt{r_1} / \sqrt{r_2} \right) = 2 \left( \sqrt{r_1 / r_2} - 1 \right) / \left( \sqrt{r_1 / r_2} + 1 \right)$ rather than the "logarithm" $\ln \sqrt{r_1 / r_2}$. Let us show it. For that let us note that although the exact function $\varphi(z)$ is continuous everywhere, its derivative $\varphi'(z)$ suffers a jump at $z = z_1$. This property follows from the exact equation integration around $z_1$. A continuous combination at $z = z_1$ is the product $\left( \sqrt{\tilde{r}} \cdot \varphi'(z) \right)$: $\sqrt{r_1} \cdot \varphi'(z_1 - 0) = \sqrt{r_2} \cdot \varphi'(z_1 + 0)$.

When we represent an exact function as a spectral sum $\varphi_n(z) = \sum C_{mn} \varphi_m^{(0)}(z)$, the latter converges. But the spectral sum for its derivative $\varphi_n' = \sum C_{mn} \varphi_m^{(0)'}$ converges very badly – it is very sensitive to the spectral terms in vicinity of $z = z_1$. It is easy to understand: the spectral addenda $\varphi_m^{(0)\prime}(z_1 - 0)$ and $\varphi_m^{(0)\prime}(z_1 + 0)$ differ infinitesimally, but their sum $\sum C_{mn} \varphi_m^{\prime}$ suffers a jump. That means the infinitesimal quantities are summed up into a finite one.

In order to correctly calculate the PT terms, let us find the perturbation operator action on the exact function $\varphi_n$ *before* the spectral decomposition of the latter (i.e., as it acts in the exact equation):

$$
\int \varphi_k^{(0)} \left( \hat{V} \varphi_n \right) dz = \int \varphi_k^{(0)} \left( \frac{1}{\tilde{r}} \right)^{\prime} \left( \sqrt{\tilde{r}} \cdot \varphi_n^{\prime} \right) dz =
$$
$$
= \left( \frac{1}{\sqrt{r_2}} - \frac{1}{\sqrt{r_1}} \right) \varphi_k^{(0)}(z_1) \left( \sqrt{r_1} \cdot \varphi_n^{\prime}(z_1 - 0) \right) = \left( \frac{1}{\sqrt{r_2}} - \frac{1}{\sqrt{r_1}} \right) \varphi_k^{(0)}(z_1) \left( \sqrt{r_2} \cdot \varphi_n^{\prime}(z_1 + 0) \right) \tag{A3.1a}
$$

As the two terms above are equal, the first term may be taken with the weight $\vartheta$ and the second one with the weight $(1 - \vartheta)$, where $\vartheta = \sqrt{r_2} / \left( \sqrt{r_1} + \sqrt{r_2} \right)$:

$$
\int \varphi_k^{(0)} \left( \hat{V} \varphi_n \right) dz = \left( \frac{1}{\sqrt{r_2}} - \frac{1}{\sqrt{r_1}} \right) \varphi_k^{(0)}(z_1) \cdot \frac{\sqrt{r_1} \cdot \sqrt{r_2}}{\sqrt{r_1} + \sqrt{r_2}} \cdot \left[ \varphi_n^{\prime}(z_1 + 0) + \varphi_n^{\prime}(z_1 - 0) \right] \tag{A3.1b}
$$

This result is still expressed via unknown exact function $\varphi_n$ that has a derivative jump. However the sum of continuous functions $\sum C_{mn} \varphi_m^{(0)\prime}(z)$ at $z = z_1$ has a quite certain (unambiguous) value equal to $\frac{1}{2} \left[ \varphi_n^{\prime}(z_1 + 0) + \varphi_n^{\prime}(z_1 - 0) \right]$. Then the matrix element (A3.1) can be written again as a *spectral sum* of the known non-perturbed eigenfunctions $\varphi_n^{(0)}$:

$$
\int \varphi_k^{(0)} \left( \hat{V} \varphi_n \right) dz = 2 \frac{\sqrt{r_1 / r_2} - 1}{\sqrt{r_1 / r_2} + 1} \sum_m C_{mn} \varphi_k^{(0)}(z_1) \varphi_m^{(0)\prime}(z_1) = \sum_m C_{mn} \overline{V}_{km}. \tag{A3.2}
$$

I call the quantities $\overline{V}_{nm} = \xi_2 \cdot \varphi_n^{(0)}(z_1) \varphi_m^{(0)\prime}(z_1)$ the **spectral analogues of the matrix elements** (compare them with (11)). They replace the "vulgar" matrix elements (11) in the correct perturbation theory.

The right small parameter is actually $\xi_2$:

$$
\xi_2 = \xi_2(\sqrt{r_1} / \sqrt{r_2}) = 2 \frac{\sqrt{r_1 / r_2} - 1}{\sqrt{r_1 / r_2} + 1}. \tag{A3.3}
$$

It differs from the "logarithm" $\ln \sqrt{r_1 / r_2}$ only starting from the third order and with a small coefficient:

$$
\xi_2 \approx \ln \sqrt{r_1 / r_2} - \frac{1}{12} \left( \ln \sqrt{r_1 / r_2} \right)^3 + \dots \quad \text{or} \quad \ln \sqrt{r_1 / r_2} \approx \xi_2 + \frac{1}{12} \xi_2^3 + \dots.
$$

It can be "easily" verified that the exact solution expansion $\lambda_0 \big|_{(z_1 = 0.5)} = \left\{ 2 \cdot \operatorname{arctg} \left[ (r_1 / r_2)^{1/4} \right] \right\}^2$ in powers of $\xi_2$ or in powers of $\varepsilon = \ln(r_1 / r_2)$, whatever, coincides with the *correct* PT expansion (it is possible to perform the third order PT correction calculation), for example:

$$\lambda_0^{(3)} = \frac{\pi^2}{4}\left(1 + \frac{\varepsilon}{\pi} + \frac{\varepsilon^2}{4\pi^2} - \frac{\varepsilon^3}{96\pi}\right), \quad \varepsilon = \ln(r_1/r_2). \tag{A3.4}$$

By the way, when $\varepsilon \approx 1/(2 \cdot 137)$, the bracket expression resembles the electron anomalous magnetic moment expansion in powers of $\alpha$. The PT series accuracy is rather high when the expansion parameter is small (see **Section 4**).

Thus, in the multi-layer problem one encounters the so called **perturbative-spectral non-commutativity** phenomenon when the perturbation operator action on the exact function (A3.1) is not equal to the sum of its actions on the spectral addenda. Without taking this phenomenon into account, one works with a "vulgar" PT, i.e., with a wrong small parameter ("logarithm"). In practical applications the difference between the correct (A.3.3) and "vulgar" PT (11) starts from the third order (see (A3.4)) and thus may be sufficiently small to worry about, but in principle the "vulgar" PT is wrong. For example, it gives the opposite sign at the third order term in (A3.4) and wrong all the higher order terms (see it on page 22).

No one perturbation operator (an "effective" operator $\hat{V}_{eff}(z)$) in a "vulgar" PT may give the matrix elements $(V_{eff})_{nm} = \overline{V}_{nm} = \xi_2 \cdot \varphi_n^{(0)}(z_1)\varphi_m^{(0)\prime}(z_1)$ with correct formula (A3.3) for $\xi_2$. However, one may safely use the "vulgar" PT series where the logarithm is replaced with $\xi_2$ because it is the sole difference between the rigorous and vulgar PT versions [1–3].

### A3.2. An iterative procedure for eigenfunctions

The original equation $\left[\dfrac{d^2}{dx^2} + \lambda \cdot r(x)\right]\psi(x) = 0$ can be cast into the form where the eigenfunction $\psi$ is "a source of itself": $\dfrac{d^2}{dx^2}\psi(x) = -\lambda \cdot r(x) \cdot \psi(x)$. Its formal solution is expressed via the Green's function $G_\lambda(x, y)$ (12): $\psi(x) = \lambda \cdot \displaystyle\int_a^b r(y) \cdot \psi(y) \cdot G_0(x, y)dy$. Formula (23c) can be understood a first iteration for the ground state eigenfunction with specifically chosen $\lambda_0$ and $\psi^{(0)}(x)$ on the right-hand side. For the discussed SLP, the approximate analytical formula (23c) is simple (here $\theta(x)$ is the Heaviside stepwise ($0 \to 1$) function):

$$\psi_0^{GF}(x) = 4\lambda_0^{GF}\frac{z_b^2}{\pi^2}\left[\varphi_0^{(0)}(z(x))\left(\frac{\theta(x_1 - x)}{\sqrt[4]{r_1}} + \frac{\theta(x - x_1)}{\sqrt[4]{r_2}}\right) + \left(\frac{1}{\sqrt[4]{r_1}} - \frac{1}{\sqrt[4]{r_2}}\right)\varphi_0^{(0)}(z(x_1)) \cdot \theta(x - x_1) + \right.$$

$$\left. + \sqrt{\frac{2}{z_b}}\frac{\pi}{2z_b}\cos\left(\frac{\pi}{2}\frac{z(x_1)}{z_b}\right)G_0(x, x_1)\left(r_2^{1/4} - r_1^{1/4}\right)\right] \tag{A3.5}$$

Numerical data show simplicity and a rather good accuracy of this approximation for all values of $x_1 \in (0,1)$ and $r_1/r_2 \in (4.5^{-1}, 4.5)$ – its relative error does not exceed $\approx \pm5\%$ within this region of these input parameters for (7) (see Fig. 15 – 18 in **Appendix 6** for the relative accuracies and Fig. 19, 20 for $\psi_0(x)$ obtained for the extreme values of $r_1/r_2$ and variable $x_1$). No additional normalization is necessary.

### A3.3. A second-order PT formula for $\lambda_n$

This formula, obtained for the boundary conditions $\psi(0) = \psi'(1) = 0$, is mainly borrowed from my preprint [1]:

$$\lambda_n^{PT(2)} = \lambda_n^{PT(1)} + \varepsilon^2 \frac{\sqrt{\lambda_n^{(0)}} \sin\left(2\sqrt{\lambda_n^{(0)}} z_1\right)}{8 z_b} \left\{ \frac{\sin\left(2\sqrt{\lambda_n^{(0)}} z_1\right)}{2\sqrt{\lambda_n^{(0)}} z_b} - \left(1 - 2\frac{z_1}{z_b}\right) \cos\left(2\sqrt{\lambda_n^{(0)}} z_1\right) \right\}, \quad (A3.6)$$

in particular, an explicit form of matrix elements (11) and formulae (4.2), (4.3) [1] were used to re-derive this result (formulae from [1] contain some minor typos). (A3.6) has a better numerical precision than $\lambda_n^{PT(1)} = \lambda_n^{PT(0)} + \bar{V}_{nn}$ at small and "moderate" values of $\varepsilon$. It confirms also my statement that the sign at $\varepsilon^2$ is not obligatorily negative when $n = 0$.

### APPENDIX 4

As I mentioned in the main text, the true small parameter in the perturbative expansion (3) – (6) is the rate of spatial change of $r(x)$. For a two-layer system (7) this rate if too big and the spectral decomposition coefficients $F_{mn}$ of a discontinuous function $\Phi_n(z) = \tilde{r}^{1/4}(z)\psi_n(x(z)) = \sum_m F_{mn}\Phi_m^{(0)}(z)$ over continuous functions $\Phi_n^{(0)}(z)$, expressed via PT series $F_{mn}^{PT} = F_{mn}^{(0)} + F_{mn}^{(1)} + F_{mn}^{(2)} + ...$ diverge starting from $F_{mn}^{(1)}$. The same statement is valid for the eigenvalues $\lambda_n^{PT} = \lambda_n^{(0)} + \lambda_n^{(1)} + \lambda_n^{(2)} ...$ since $U_{nn}$ and $U_{nm}$ diverge. One cannot advance father than the zeroth-order approximations. A natural conclusion from this fact is obviously to try from the very beginning to reformulate the problem (1) or (3) in terms of better initial approximations with a really small parameter, as in **Section 2**, for example.

However one **can** obtain finite series for $\Phi_n(z)$ and $\lambda_n$ (2) – (3) from PT series (4) – (6) if one manages to sum up (or, loosely speaking, to "hide up") the divergent terms in new, "rebuilt" or "dressed" eigenfunctions and eigenvalues. Let us show this for a methodological reason. For that and to simplify our demonstration let us apply the following boundary conditions: $\psi(a) = \psi(b) = 0$. Then the zeroth-order approximations for $\Phi_n(z)$ and $\varphi_n(z)$ coincide for any $r(x)$ dependence: $\Phi_n^{(0)}(z) \equiv \varphi_n^{(0)}(z)$ (no shifts of $\alpha_{a,b}$ due to $\tilde{r}'\big|_{z=a,b} \neq 0$ arise).

The PT series (4) – (6) with $U(z)$ are divergent. They can be transformed into (rewritten as) convergent PT series like

$$\Phi_n^{PT} = \chi_N + \sum_{M \neq N} \frac{(\delta U)_{MN}}{\Lambda_N - \Lambda_M} \chi_M + ..., \; N = n, \; M = m, \quad (A4.1)$$

$$\lambda_n^{PT} = \Lambda_N + (\delta U)_{NN} + \sum_{M \neq N} \frac{(\delta U)_{MN} \cdot (\delta U)_{NM}}{\Lambda_N - \Lambda_M} + ..., \quad (A4.2)$$

where $(\delta U)_{MN}$ are finite matrix elements of some operator $\left(\widehat{\delta U}\right) = U - \hat{U}^R$ calculated in the basis of new eigenfunctions $\{\chi_N\}$ satisfying the equation (see **Appendix 5**):

$$\left[\frac{d^2}{dz^2} + \Lambda_N - \hat{U}^R(z)\right]\chi_N(z) = 0, \tag{A4.3}$$

$$(\delta U)_{MN} = \left(\chi_M, \left(\widehat{\delta U}\right)\chi_N\right) \tag{A4.4}$$

Although numerically $N = n$, $M = m$, I write the new basis subscripts in capital letters to distinguish the matrix elements calculated in different basises (see the right definition of the scalar product $(\chi_N, \chi_M)$ in **Appendix 5**).

With appropriately choosing the basis functions $\chi_N$ or, which is the same, the operator $\hat{U}^R$, one can make the matrix elements (A4.4) be finite whereas the PT series for $\chi_N$, $\Lambda_N$, and $(\delta U)_{MN}$ in powers of $\hat{U}^R$ in the basis $\{\varphi_n^{(0)}\}$ can themselves be divergent due to divergence of the matrix elements $U^R_{mn}$:

$$\chi_N^{PT} = \varphi_n^{(0)} + \sum_{m \neq n}\frac{U^R_{mn}}{\lambda_n^{(0)} - \lambda_m^{(0)}}\varphi_m^{(0)} + ...,$$

$$\Lambda_N^{PT} = \lambda_n^{(0)} + U^R_{nn} + \sum_{m \neq n}\frac{U^R_{mn}U^R_{nm}}{\lambda_n^{(0)} - \lambda_m^{(0)}} + ..., \ n = N. \tag{A4.5}$$

Injecting the divergent series (A4.5) in formulate (A4.1), (A4.2), *and* (A4.4) gives the original divergent series (4) – (6) where $U_{mn} = U^R_{mn} + (\delta U)_{mn}$ (see **Appendix 5**).

The process of grouping divergent correction into series of $\chi_N^{PT}$ and $\Lambda_N^{PT}$ and replacing the latter with their exact (finite) expressions $\chi_N$ and $\Lambda_N$ can be called "rebuilding" or "exact dressing" the eigenfunctions due to the perturbation $\hat{U}^R$ (the superscript "R" stands for "Rebuilding"). Then the series (A4.1) and (A4.2) may be called "rebuilt" PT series. This boils down to choosing a better basis for the spectral decomposition of $\Phi(z)$.

In general case the clue allowing choosing a new basis $\{\chi_N\}$ (or the operator $\hat{U}^R$) should follow form physical or/and mathematical properties of the exact solutions $\Phi_n$ (or $\psi_n$), **if the latter is known to exist**. In our case of two-layer system this clue is elementary (because we know the exact answer): as far as the exact function $\Phi(z)$ has a jump causing the matrix element divergences, then why not to decompose $\Phi(z)$ over some eigenfunctions having a similar jump, for example, over $\chi_N = \tilde{r}^{1/4}\varphi_n^{(0)}$? For the given choice of $\chi_N$ the eigenvalues $\Lambda_N$ coincide with $\lambda_n^{(0)}$ and the operator $\hat{U}^R$ is equal to:

$$\hat{U}^R(z) = U(z) - \frac{1}{2}\left(\ln\sqrt{\tilde{r}}\right)'^2 + \left(\ln\sqrt{\tilde{r}}\right)'\frac{d}{dz}, \tag{A4.6}$$

The matrix elements of the remaining operator $\left(\widehat{\delta U}\right)(z) = U(z) - \hat{U}^R(z) = -2\left(\tilde{r}^{1/4}\right)'\frac{d}{dz}\left(\frac{1}{\tilde{r}^{1/4}}\right)$ in the basis $\{\chi_N\}$ are finite and are equal to $(\delta U)_{NM} = (\delta U)_{nm}^{(0)} = \varphi_n^{(0)}(z_1)\cdot\varphi_m^{(0)'}(z_1)\cdot\ln\sqrt{r_1/r_2}$, i.e., coincide with formula (11). And the solution for $\psi_n^{PT} = r^{-1/4}\Phi_n^{PT} = r^{-1/4}\sum_M\left(F^R_{MN}\right)^{PT}\chi_M = \sum_m\left(F_{mn}\right)^{PT}\varphi_m^{(0)}$ coincides with the "vulgar" PT solution from (9) – (10). (In order to obtain the correct PT with small parameter $\xi_2$, one has to carefully take into account the "perturbative-spectral non-commutativity" phenomenon: $\int\bar{\chi}_K\left(\left(\widehat{\delta U}\right)\Phi_N\right)dz \neq \int\sum_M F^R_{NM}\left(\bar{\chi}_K\left(\widehat{\delta U}\right)\chi_M\right)dz$, see **Appendices 3** and **5** for explanations).

I will not develop right here the detailed proofs of these statements and provide the correct rather than "vulgar" PT consideration in such a "re-summation" approach (see **Appendix 5**). The main statement here is that the divergent corrections *can* in principle be summed up exactly into new basis eigenfunctions and the finite PT series can thus be obtained from divergent ones, i.e., in the frame of the problem formulation with a very singular perturbation (3). However it only is possible in case of existence of the original equation physical solutions.

What is interesting here is that whatever dependence of $r(x)$ is, there is an identity: $\Lambda_N \equiv \lambda_n^{(0)}$. On the other hand the zeroth-order approximation $\Lambda_N^{(0)}$ from (A4.5) is also equals to $\lambda_n^{(0)}$. That means the whole divergent series "tail" from (A4.5): $U_{nn}^R + \sum_{m \neq n} \dfrac{U_{mn}^R U_{nm}^R}{\lambda_n^{(0)} - \lambda_m^{(0)}}...$, having been summed up, vanishes. It is equal identically to zero although in each perturbative order it diverges. I called such useless expansions "blank". This surprising property is finally explained with the fact that all perturbative corrections here, starting from the first order, represent a **difference** between some function and its perturbation series. (See the proof of this fact in **Appendix 5**). Factually this is a rigorous mathematical explanation why *discarding* the carefully separated divergent corrections in each PT order may give good finite series – summed up, such divergent corrections result in zero anyway.

The same statement is valid for the matrix elements $\left( \delta U \right)_{MN}^{PT}$ expansion in powers of $\hat{U}^R$:
$\left( \delta U \right)_{MN}^{PT} = \left( \delta U \right)_{mn}^{(0)} + \left( \delta U \right)_{mn}^{(1)} + \left( \delta U \right)_{mn}^{(2)} + ....$ All corrections to $\left( \delta U \right)_{mn}^{(0)}$, having been properly (formally) summed up, cancel ( see the proof of these facts in **Appendix 5**, formulae (A5.12), (A5.13)).

Replacing the divergent series $\chi_N^{PT}$ and $\Lambda_N^{PT}$ with their exact expressions $\chi_N$ and $\Lambda_N$ only is possible if we know *how* to obtain the exact expressions. In this case the basis "rebuilding" or the eigenfunction "exact dressing" is as legitimate as any other mathematical operation even though it may need a special (non-linear, Borel, etc.) summation of badly convergent series. But knowing the exact expressions or how to sum up certain terms into finite functions is an extremely rare case. In QED, for example, one does not know how to and into what sum up divergent corrections. It was an obstacle for PT calculations for about 20 years. Finally, the divergence *discarding* or "renormalization" prescription was worked out as a way of obtaining finite solutions.

In our case (1) such a discarding means "another way" of obtaining *finite* expressions for $\Phi_n^{PT}$ and $\lambda_n^{PT}$ from divergent series (4) – (6). Such a discarding, without knowing the exact solutions, the origin of (3), and the exact relationships (2), may be somewhat "justified" with comparison, for example, of $\lambda_0^{PT}$ with the experimental data on the regular regime of heat conduction (or diffusion) that is determined only by the slowest decaying exponential $e^{-\lambda_0 t}$, $t \geq 1/\lambda_0$. If we choose as the "renormalizing" condition the relationship $\bar{U}_{00} = \Phi_0^{(0)}(z_1) \cdot \Phi_0^{(0)\prime}(z_1) \cdot \ln \sqrt{r_1/r_2}$ (see formula (8)), and then, by analogy with it $\bar{U}_{nm} = \Phi_n^{(0)}(z_1) \cdot \Phi_m^{(0)\prime}(z_1) \cdot \ln \sqrt{r_1/r_2}$ , i.e., if we remove (subtract) the operator $U + \left( \ln \sqrt{r} \right)' d/dz$ from equation (3) and only leave $\bar{U} = -\left( \ln \sqrt{r} \right)' d/dz$ (kind of adding an exact "counter-term" to the original equation), then we obtain the correct (correct up to the second order) series (at least for $\lambda_n^{PT}$ ) and a good numerical agreement, especially if the formally small parameter $\ln \sqrt{r_1/r_2}$ is small indeed (see **Section 4**). Isn't a success of "subtraction prescription"? If we believe so, then we are bound to deviate from the correct expansion in the third order due to perturbative-spectral non-commutativity phenomenon in our particular case.

If we analyse more carefully the *so obtained new equation*, we may come to the correct PT for $\lambda_n$ (i.e., with correct $\xi_2$ ) since it follows from the exact equation properties, but we will suppose that we cannot do it as we cannot do it in general case of QFT (one makes subtractions perturbatively, without knowing the exact solution properties). So, such a discarding, "justified" solely by good agreement with some "experimental" or numerical data ($(\lambda_0)_{\text{exp.}}$ or $(\lambda_0)_{\text{numerical}}$ from the transcendental equation), is clearly not legitimate mathematical action. It is counting on luck and miracles. Numerical "success" in one one case gives a bad example for following in other cases. That is why one encounters non-renormalizable theories – one counts on lucky guess (by analogy) rather than on physically and mathematically meaningful formulations, and one's naive "prescription" fails.

In our example, as a result of discarding, we lose the factor $\tilde{r}^{1/4}(z)$ in the exact relationship between $\psi(x)$ and $\Phi(z)$. This means "sacrificing" some part of the initial "potential" $U(z)$, i.e., working with *another* Hamiltonian (in our case, it is roughly (10) instead of (3)). In other words, *discarding (or adding the counter-terms) means postulating another equation for the phenomenon description*. Although discarding (or adding a counter-term) removes the divergences, it does not automatically mean that the new, finite series correspond to the right solutions. In our example we obviously obtain the (vulgar) PT solution $\varphi_n^{PT}$ for $\Phi_n^{PT}$, i.e., without factor $r^{1/4}(z)$ which is only "acceptable" in case if we do not know (or are not going to use) the exact relationship (2) and which is only "good" up to the second order. Starting from the third PT order, the series with $\ln\left(\sqrt{r_1/r_2}\right)$ (i.e., in vulgar PT) are wrong.

Thus, it is especially incorrect to "*determine*" the numerical value of the small parameter $\varepsilon$ (a charge or coupling constant in QED) by comparing such a dubiously obtained solutions (wrong in our case due to wrong $\bar{U}_{nm}$), for example, by comparing $\left(\lambda_0^{(3)}\right)_{\text{"vulgar"}} = \frac{\pi^2}{4}\left(1 + \frac{\varepsilon}{\pi} + \frac{\varepsilon^2}{4\pi^2} + \frac{\varepsilon^3}{96\pi}\right)$ with the "experimental or numerical data" (A4.3, Fig. 6) (kind of the "renormalized" coupling constant "fitting"):

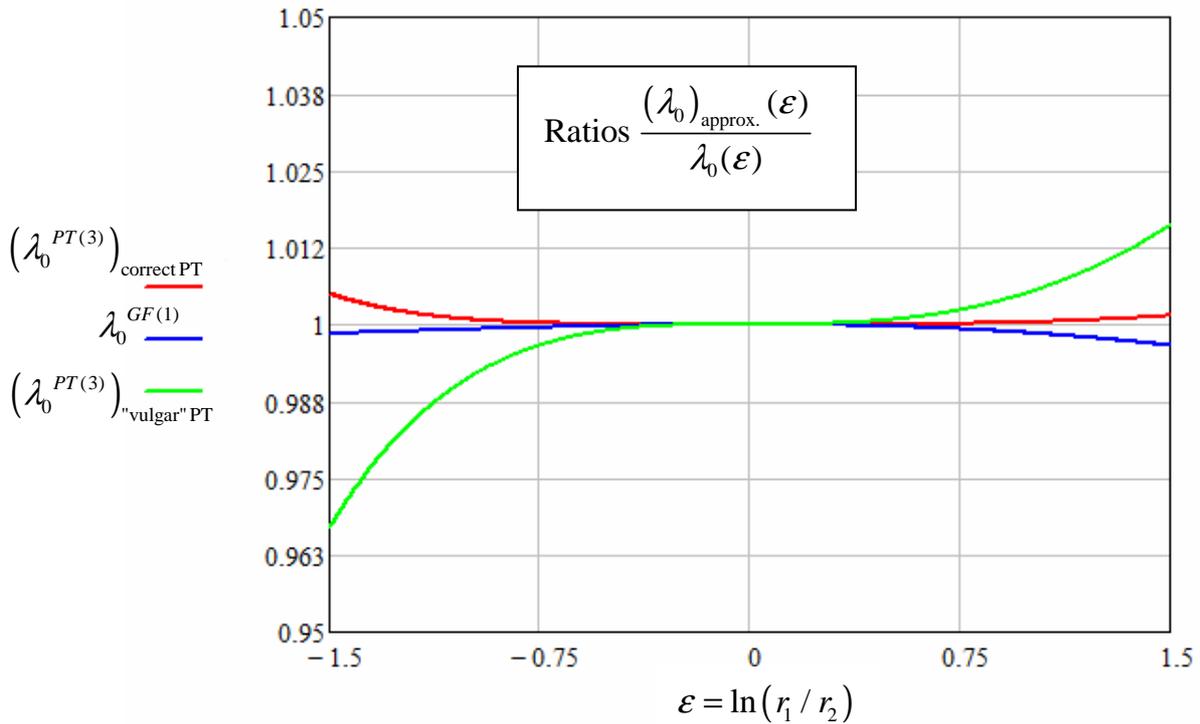

Fig. 14. Ratios of approximate values of $\lambda_0$ from (A3.4), from (18) (Fig. 6) and "vulgar" one given above to the exact eigenvalue $\lambda_0(\varepsilon)\big|_{(z_1=0.5)}$ from **Appendix 3**.

Perturbatively changing the basis (or the eigenfunction "dressing" or "rebuilding") is a quite painful operation even in case when one knows the exact result and one is sure to be able to carry it out *properly*. It is much more natural and easier to start directly from better initial approximations (better basis) (9) – (10) and obtain the finite series from the very beginning. The PT consideration (9) – (10), (A3.2) – (A3.3) of the problem

(1) with (7) is a **short-cut** to the correct results whereas the perturbative rebuilding the basic functions (A4.1) – (A4.4) is a long and hazardous way.

This article has been conceived with the purpose to encourage the researchers to seek physical and/or mathematical "short-cuts" or reformulations in their divergence problems rather than seek "justifications" to discarding (subtraction) prescriptions.

## APPENDIX 5

### A5.1. Exact summation of divergent corrections

Let us present the "perturbation" operator $U(z)$ in (3) as a sum $U = \hat{U}^R + \left(\widehat{\delta U}\right)$ where

$$\hat{U}^R(z) = U(z) - \frac{1}{2}\left(\ln\sqrt{\tilde{r}}\right)'^2 + \left(\ln\sqrt{\tilde{r}}\right)'\frac{d}{dz}, \qquad \left(\widehat{\delta U}\right)(z) = -2\left(\tilde{r}^{1/4}\right)'\frac{d}{dz}\left(\frac{1}{\tilde{r}^{1/4}}\right) \qquad (A5.1)$$

$$\left[\frac{d^2}{dz^2} + \lambda_n - \hat{U}^R(z) - \left(\widehat{\delta U}\right)(z)\right]\Phi_n = 0 \qquad (A5.2)$$

The mixed boundary conditions (1) are transformed into:

$$\left[\alpha_{a,b} + \frac{1}{2}\sqrt{\tilde{r}}'(z_{a,b})\right]\Phi(z_{a,b}) \mp \sqrt{\tilde{r}(z_{a,b})}\cdot\Phi'(z_{a,b}) = 0. \qquad (A5.3)$$

They are different from (19) if the derivatives of $\tilde{r}(z)$ are not equal to zero at $z_a$ and $z_b$. For simplicity we will consider those cases where the values $r'|_{a,b}$ do not play any role, for example, when $r'|_{a,b} = 0$ or $\Phi(z_{a,b}) = 0$, whatever. Then the zeroth-order (initial) approximations of $\Phi_n(z)$ and $\varphi_n(z)$ coincide: $\Phi_n^{(0)}(z) \equiv \varphi_n^{(0)}(z)$.

Now, let us consider the solutions $\chi_N$, $\Lambda_N$ of equation (A4.3). They define a new basis $\{\chi_N\}$ with the following scalar product: $(\chi_N, \chi_M) = \int \frac{1}{\sqrt{\tilde{r}(z)}}\chi_N(z)\chi_M(z)dz$. The necessity of the weight $\tilde{r}^{-1/2}$ in it is evident from definition of $\chi_N$ via $\varphi_n^{(0)}$ as well as from equation (A4.3).

We may define the functions $\bar{\chi}_N = \frac{\chi_N}{\sqrt{\tilde{r}}} = \frac{\varphi_n^{(0)}}{\sqrt[4]{\tilde{r}}}$ as "conjugated" to the functions $\chi_N$ in the sense of *weightless* scalar product: $(\bar{\chi}_N, \chi_M) = \int \bar{\chi}_N\chi_M dz = \delta_{NM}$. The functions $\bar{\chi}_N$ obey the same equation (A4.3) where $\tilde{r}(z)$ is just replaced with $\tilde{r}^{-1}(z)$. The perturbation series for $\bar{\chi}_N$ can be obtained from the series $\chi_N^{PT}$ just by changing $U^R_{mn} \to U^R_{nm}$ in the series of $\chi_N^{PT}$. It is useful since it is namely $\bar{\chi}_N$ that should be expanded in PT series in powers of $\hat{U}^R$ in the matrix element $(\delta U)_{MN}^{PT} = \int \bar{\chi}_M^{PT}\left(\widehat{\delta U}\right)\chi_N^{PT}dz$ perturbative expansion.

Let us find the perturbation operator $\left(\widehat{\delta U}\right)$ action on the exact function $\Phi_n$ *before* the spectral decomposition (as it acts in the exact equation):

$$\int \bar{\chi}_K \left( \widehat{\delta U} \right) \Phi_n \, dz = -2 \int \bar{\chi}_K \left( \tilde{r}^{1/4} \right)' \frac{1}{\sqrt{\tilde{r}}} \left( \sqrt{\tilde{r}} \cdot \varphi_n' \right) dz = \int \varphi_k^{(0)} \left( \frac{1}{\sqrt{\tilde{r}}} \right)' \left( \sqrt{\tilde{r}} \cdot \varphi_n' \right) dz =$$

$$= \sum_m C_{mn} \left[ \xi_2 \cdot \varphi_k^{(0)}(z_1) \cdot \varphi_m^{(0)\prime}(z_1) \right] = \sum_m C_{mn} \bar{V}_{km}, \qquad (A5.4)$$

which coincides with the expressions (A3.1) – (A3.2) from **Appendix 3**. In (A5.4) I used the exact relationships: $\Phi_n = \tilde{r}^{1/4} \varphi_n$, $K = k$, and $\bar{\chi}_K = \tilde{r}^{-1/4} \varphi_k^{(0)}$. Thus the correct "matrix elements" $\overline{\left( \delta U \right)}_{NM}$ to be used in (A4.1) and (A4.2) contain the true small parameter $\xi_2 = 2 \left( \sqrt{r_1 / r_2} - 1 \right) \big/ \left( \sqrt{r_1 / r_2} + 1 \right)$ rather than the "logarithm" $\ln \sqrt{r_1 / r_2}$ :

$$\overline{\left( \delta U \right)}_{NM} = \xi_2 \cdot \varphi_n^{(0)}(z_1) \cdot \varphi_m^{(0)\prime}(z_1) = \bar{V}_{nm} . \qquad (A5.5)$$

**Thus with (A5.5) we obtain the finite PT series (A4.1) and (A4.2) coinciding factually with those obtained previously for $\varphi_n$ and $\lambda_n$ from (10).**

On the other hand, the matrix elements $\overline{\left( \delta U \right)}_{NM}$ (A4.4), as well as $\chi_N$ and $\Lambda_N$ (A4.5)) "contain" divergences if expanded in powers of $\hat{U}^R$.

Let us first consider the eigenfunction expansions:

$$\chi_N = D_n \left( \varphi_n^{(0)} + \sum_{m \neq n} D_{mn} \varphi_m^{(0)} \right), \quad \bar{\chi}_N = \bar{D}_n \left( \varphi_n^{(0)} + \sum_{m \neq n} \bar{D}_{mn} \varphi_m^{(0)} \right). \qquad (A5.6)$$

The spectral coefficients $D_{mn}$ expanded up to the second order have the following expressions:

$$D_{mn}^{PT(2)} = \frac{U_{mn}^R}{\lambda_n^{(0)} - \lambda_m^{(0)}} - \frac{U_{nn}^R U_{mn}^R}{\left( \lambda_n^{(0)} - \lambda_m^{(0)} \right)^2} + \frac{1}{\lambda_n^{(0)} - \lambda_m^{(0)}} \sum_{k \neq n} \frac{U_{mk}^R U_{kn}^R}{\lambda_n^{(0)} - \lambda_m^{(0)}} , \qquad (A5.7)$$

and the coefficients $\bar{D}_{mn}^{PT(2)}$ are obtained from (A5.7) by replacing the matrix elements $U_{ik}^R$ by $U_{ki}^R$ (since it is the way how the matrix elements change upon replacing $r$ with $r^{-1}$). The PT expansions of coefficients $D_{mn}$ diverge since the matrix elements $U_{mn}^R$ diverge.

The product of the coefficients $D_n$ and $\bar{D}_n$ is determined with the normalization condition:

$$D_n \bar{D}_n = \left( 1 + \sum_{m \neq n} D_{mn} \bar{D}_{mn} \right)^{-1} . \qquad (A5.8)$$

The coefficients $D_n$ and $\bar{D}_n$ are present in all expressions as this product so there is no need to expand them separately.

Injecting these expansions into the matrix element definition $\overline{\left( \delta U \right)}_{MN}$ via $\chi$ and $\bar{\chi}$ and then, together with $\Lambda_N^{PT(2)}$ from (A4.5), into (A4.2), we obtain:

$$\lambda_n{}^{PT(2)} = \lambda_n{}^{(0)} + \left(\hat{U}^R + \left(\widehat{\delta U}\right)\right)_{nn} + \sum_{m \neq n} \frac{\left(\hat{U}^R + \left(\widehat{\delta U}\right)\right)_{mn} \cdot \left(\hat{U}^R + \left(\widehat{\delta U}\right)\right)_{nm}}{\lambda_n{}^{(0)} - \lambda_m{}^{(0)}}, \tag{A5.9}$$

that coincides with the second order formula (5) with $U = \hat{U}^R + \left(\widehat{\delta U}\right)$.

As the spectral decomposition of $\chi_N(z)$ serves solely to describe the function jump at $z = z_1$, for example:

$$\chi_0(z) = D_0 \left( 1 + \sum_{m \geq 1} D_{m0} \frac{\varphi_m{}^{(0)}(z)}{\varphi_0{}^{(0)}(z)} \right) \cdot \varphi_0{}^{(0)}(z) = \left[ \tilde{r}(z) \right]^{1/4} \cdot \varphi_0{}^{(0)}(z), \tag{A5.10}$$

the expansion of $D_{mn}$ in powers of $\left( \Delta r / \Delta z \right)$ diverges. Indeed, the step-wise factor $\left[ \tilde{r}(z) \right]^{1/4}$ expansion can be modelled with help of the Heaviside function expansion, for example:

$$\left[ \tilde{r}(z) \right]^{1/4} = \left\{ r_1 + \frac{r_2 - r_1}{1 + \exp\left[ (z_1 - z) / \Delta z \right]} \right\}^{1/4} \approx \left( \frac{r_1 + r_2}{2} \right)^{1/4} \left[ 1 + \frac{1}{8} \cdot \frac{r_1 - r_2}{r_1 + r_2} \cdot \frac{(z_1 - z)}{\Delta z} + ... \right] \tag{A5.11}$$

At any finite values of $\Delta r = \left( r_1 - r_2 \right)$ and of $(z_1 - z)$ the series (A5.11) diverges in the limit $\Delta z \to 0$ since the dimensionless distance $(z_1 - z) / \Delta z$ becomes infinitely "far" from the expansion point $z_1$. As I said in Section 1.1, such an expansion is an attempt to calculate $f(\xi_1 \to \infty, \xi_2)$ from its Taylor-Maclaurin series $f(\xi_1, \xi_2) \approx f(0, \xi_2) + f'(0, \xi_2) \cdot \xi_1 + ...$ obtained for formally small $\xi_1$. This is the true reason of divergence of the PT series (4).

Strictly speaking, these divergences cannot be discarded or "absorbed" in some constants without harm to the good sense. They can be summed up into right finite functions if *all* terms are taken into account. Attempts to carry out a *selective summing up* (for example, only the "most divergent" terms in (4) or/and (5)) may lead to absurd (non physical) results like the Landau pole. Attempts to consider some phenomenological parameters like the heat conductivities, the system width, etc., as "bare" and cut-off dependent ones are not convincing either.

### A5.2. "Blank expansions"

The divergent corrections in expansion of $\overline{\left( \delta U \right)}_{MN}$ and $\Lambda_N$ cancel if summed up properly:

$$\overline{\left( \delta U \right)}_{MN}{}^{PT} = \Phi_m{}^{(0)}(z_1) \Phi_n{}^{(0)}{}'(z_1) \xi_2 + \left\{ \int \Phi_m{}^{(0)} \left[ -\left( \ln \sqrt{\tilde{r}} \right)' \frac{d}{dz} \right] \Phi_n{}^{(0)} dz - \int \overline{\chi}_M \left( \widehat{\delta U} \right) \chi_N dz \right\}, \tag{A5.12}$$

$$\Lambda_N{}^{PT} = \lambda_n{}^{(0)} + \left\{ \int \Phi_n{}^{(0)} \frac{d^2}{dz^2} \Phi_n{}^{(0)} dz - \int \overline{\chi}_N \left( \frac{d^2}{dz^2} - \hat{U}^{(R)} \right) \chi_N dz \right\}. \tag{A5.13}$$

The figured bracket $\{...\}$ in each expression above is identically equal to zero $\{...\} = 0$ since it is a difference between two representations of the same integral. On the other hand, if $\overline{\chi}$ and $\chi$ in the integrals are expanded in powers of $\hat{U}^R$ (see (A4.5)), each figured bracket $\{...\}$ represents a divergent, but a useless perturbation series (a "blank expansion"). To prove it, here one may use the "possibility" of working with non-perturbed eigenfunctions $\Phi_n{}^{(0)}$ followed by replacing $\ln \sqrt{r_1 / r_2} \to \xi_2$ in "vulgar" matrix elements in order to obtain the correct PT results.

The divergences may also be bypassed with the problem reformulation. The latter way is preferable since reliable. We have to carefully analyze the exact original equation, derive as much as possible from it without the perturbation theory, for example, the function jump and different left and right derivatives, etc., and then try to better model the searched solutions. The "renormalization" success may serve as a hint that the problem reformulation (bypassing the divergences) can be carried out exactly instead of perturbatively.

**APPENDIX 6  (Tests of formula (A3.5))**

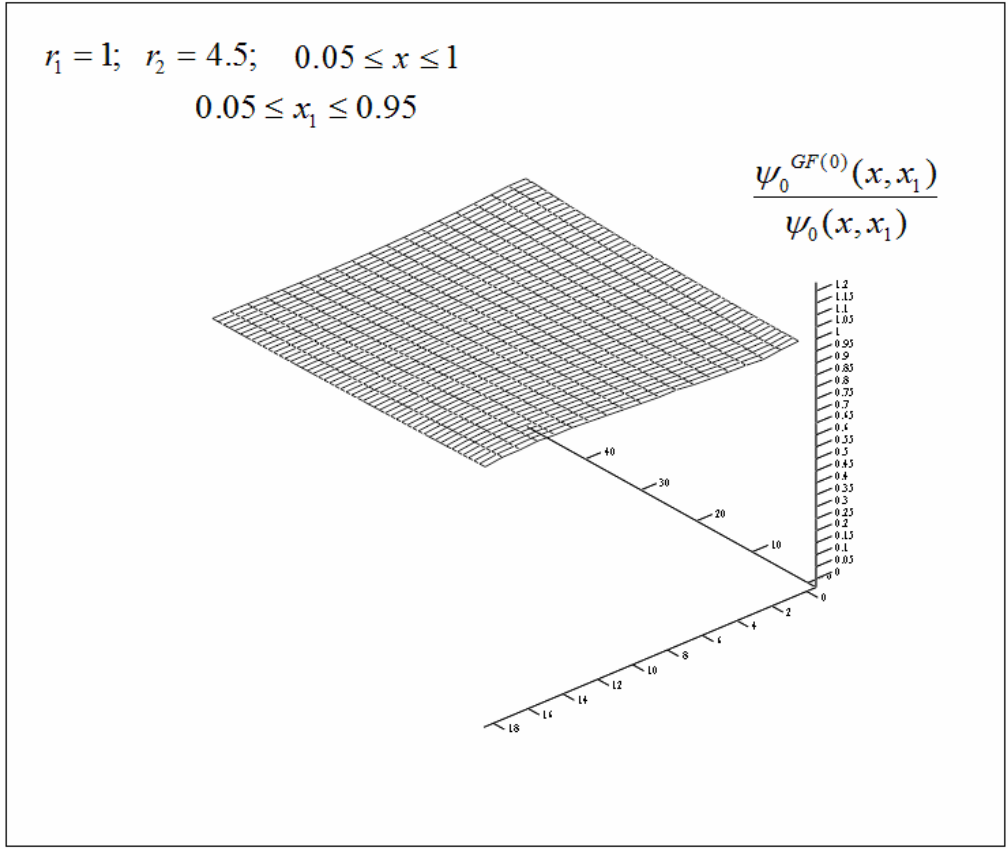

Fig. 15. Ratios of the approximate (23c) and the exact eigenfunctions at different values of $x_1$.

(19 different points of $x_1$ and 50 points along $x$. The points $x = 0$ were excluded to avoid the ratios $0/0$.)

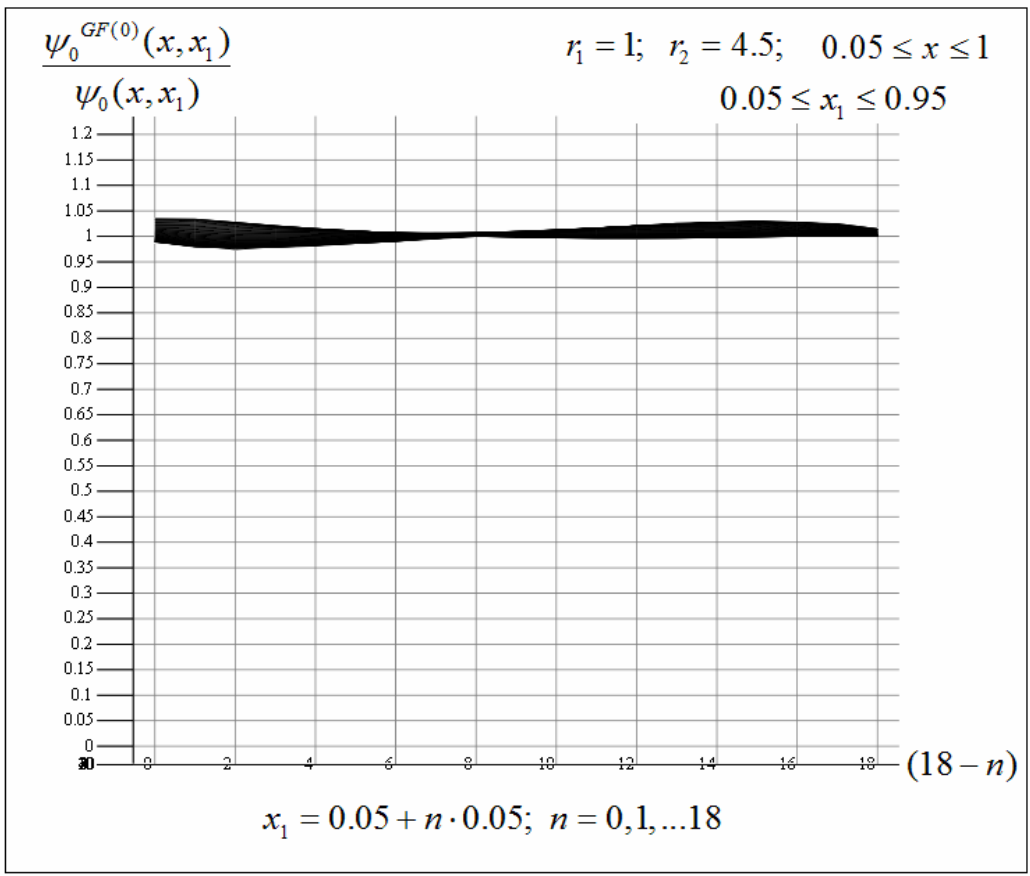

Fig. 16. Horizontal view of the ratios of the approximate (23c) and the exact eigenfunctions at different values of $x_1$.

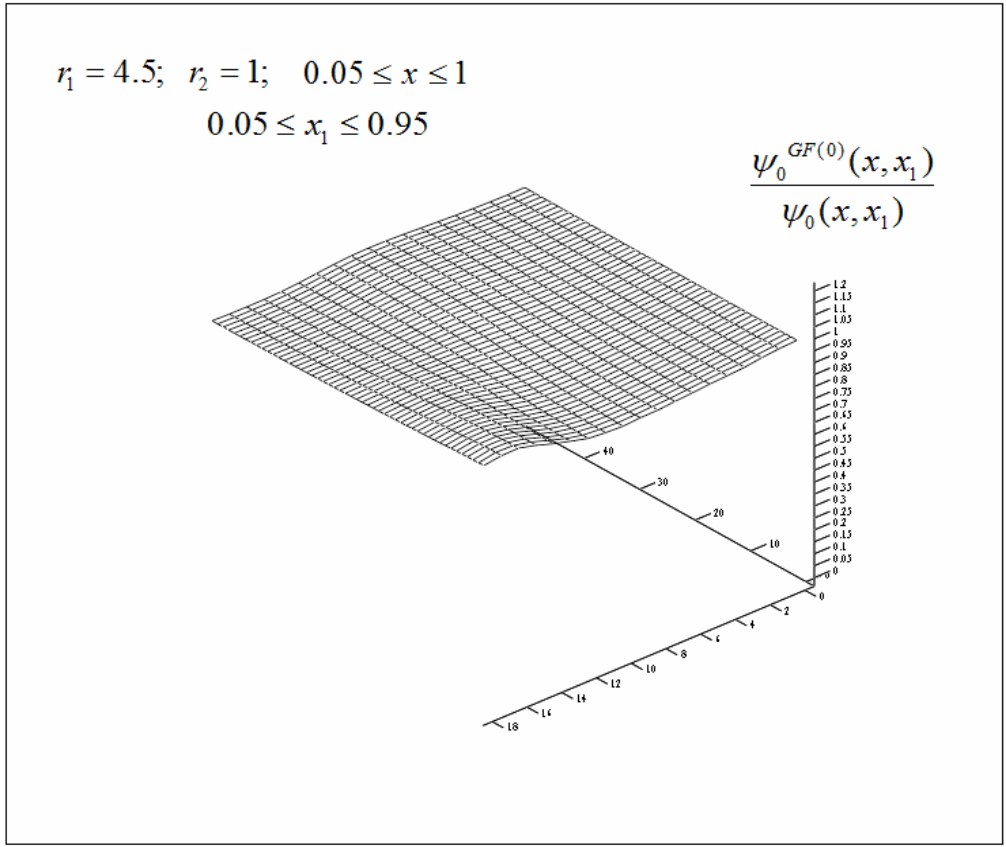

Fig. 17. Ratios of the approximate (23c) and the exact eigenfunctions at different values of $x_1$.

(19 different points of $x_1$ and 50 points along $x$. The points $x = 0$ were excluded to avoid the ratios $0/0$.)

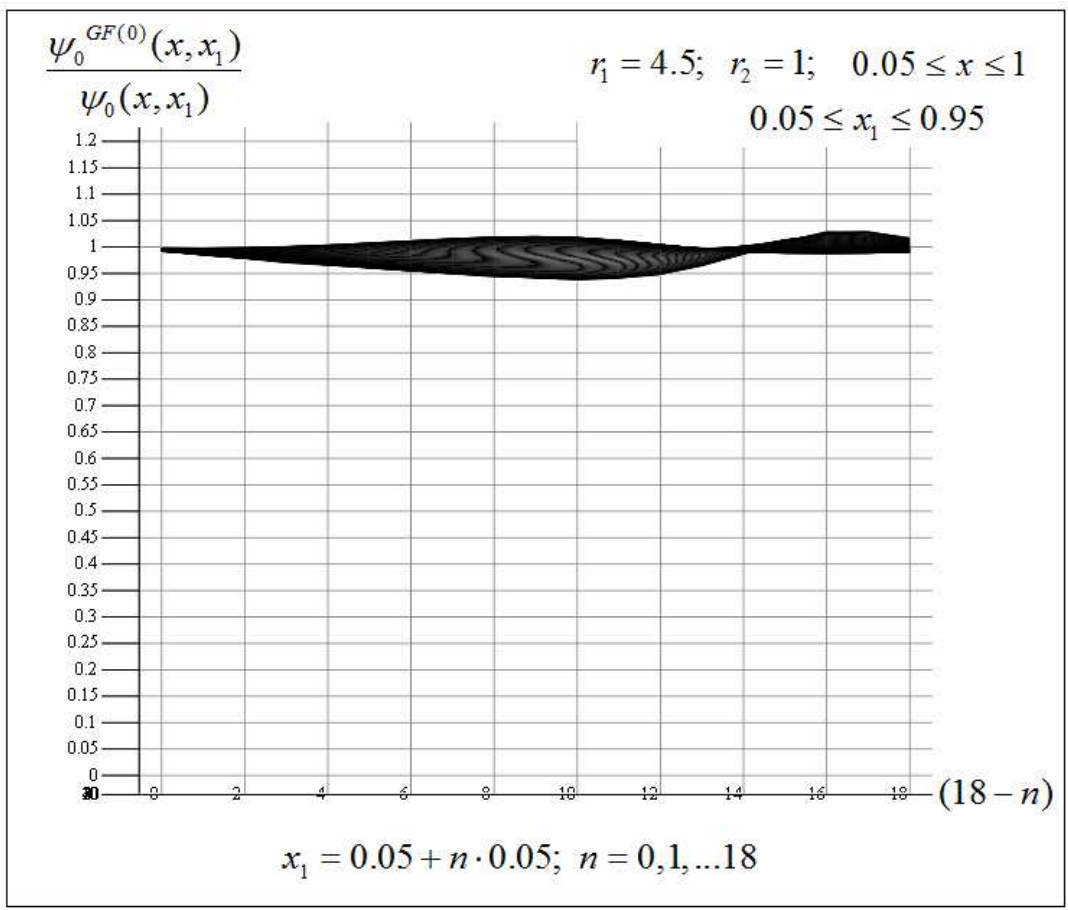

Fig. 18. Horizontal view of the ratios of the approximate (23c) and the exact eigenfunctions at different values of $x_1$.

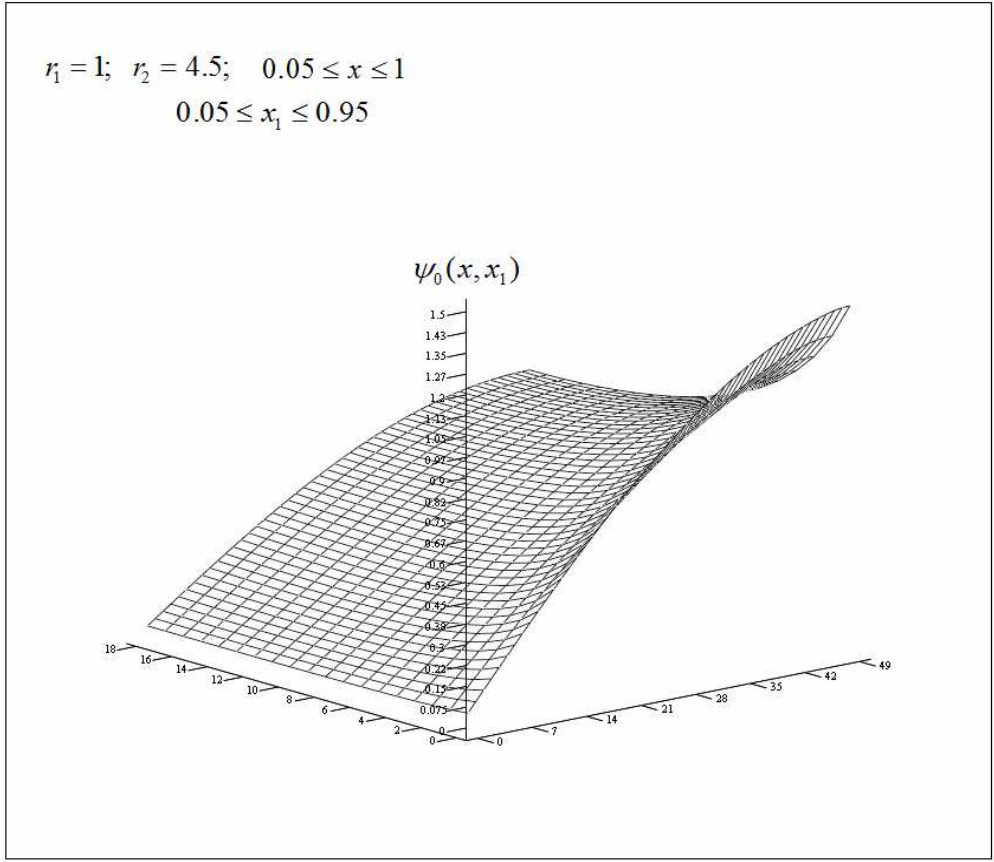

Fig. 19. Surface the exact eigenfunction $\psi_0(x)$ at different values of $x_1$ and $x$.

(19 different points of $x_1$ and 50 points along $x$. The points $x = 0$ were excluded to avoid the ratios $0/0$.)

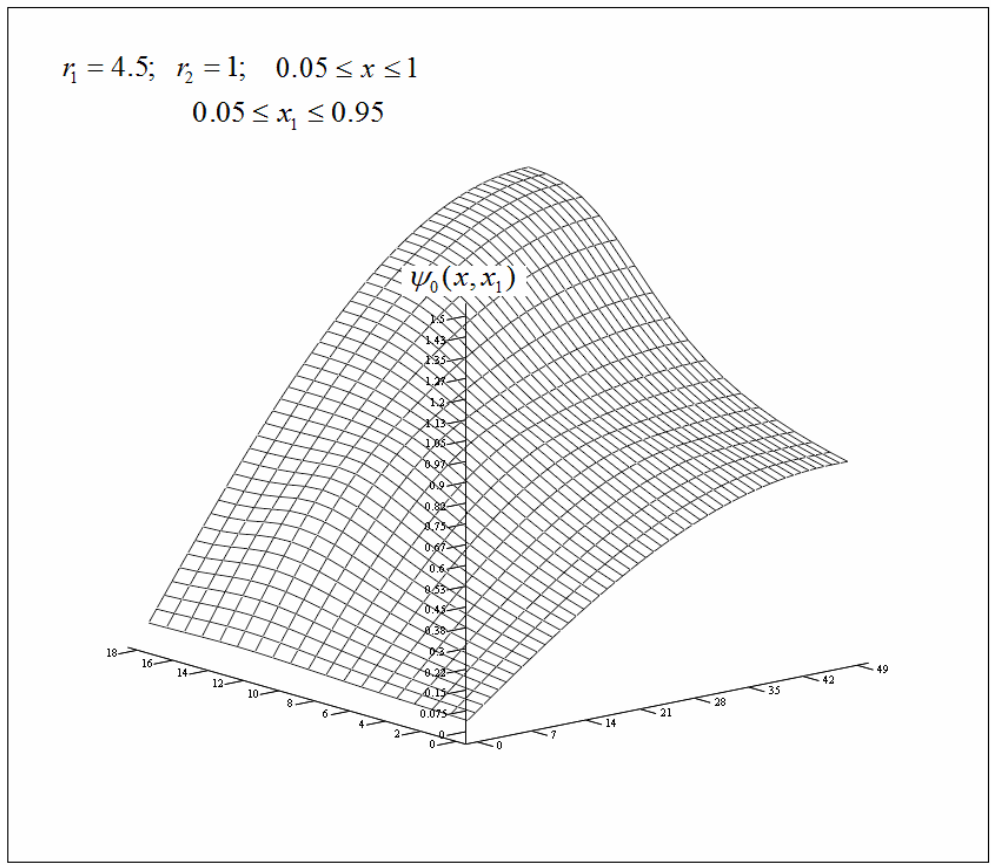

Fig. 20. Surface the exact eigenfunction $\psi_0(x)$ at different values of $x_1$ and $x$.

(19 different points of $x_1$ and 50 points along $x$. The points $x = 0$ were excluded to avoid the ratios $0/0$.)